\documentclass[10pt,journal,compsoc]{IEEEtran}

\ifCLASSOPTIONcompsoc
  \usepackage[nocompress]{cite}
\else
  \usepackage{cite}
\fi
\ifCLASSOPTIONcompsoc
  \usepackage[nocompress]{cite}
\else
  \usepackage{cite}
\fi

\usepackage{graphicx}	
\usepackage{algorithm}
\usepackage{multirow}
\usepackage{hyperref}
\hypersetup{
    colorlinks=false,
    pdfborder={0 0 0},
}
\usepackage{paralist}
\usepackage[normalem]{ulem}
\useunder{\uline}{\ul}{}
\usepackage[noend]{algpseudocode}
\usepackage{setspace}
\usepackage{nccmath}
\usepackage{booktabs} 
\usepackage[noend]{algpseudocode}
\usepackage{color}
\usepackage{xcolor}
\usepackage{amssymb}
\usepackage{amsmath}
\usepackage{amsfonts}
\usepackage{float}
\usepackage{multirow}

\usepackage{amsthm}
\usepackage[justification=centering]{caption}
\usepackage{diagbox}
\usepackage{xcolor}
\usepackage{colortbl}
\usepackage{subcaption}
\usepackage[justification=centering]{caption}
\usepackage[shortlabels]{enumitem}

\setenumerate[1]{itemsep=0pt,partopsep=0pt,parsep=\parskip,topsep=0pt}
\setitemize[1]{itemsep=0pt,partopsep=0pt,parsep=\parskip,topsep=0pt}
\setdescription{itemsep=0pt,partopsep=0pt,parsep=\parskip,topsep=0pt}

\newcommand{\best}[1]{\textbf{#1}}
\newcommand{\secBest}[1]{\uline{#1}}
\newcommand\ExpCaption[1]{%
     \captionsetup{font=small}%
     \caption{#1}}

\newcommand{\tabincell}[2]{\begin{tabular}{@{}#1@{}}#2\end{tabular}}

\newtheorem{definition}{\bf Definition}
\newtheorem{problem}{\bf Problem}
\newtheorem{example}{\bf Example}

\usepackage[switch]{lineno}
\hypersetup{
    colorlinks=false,
    pdfborder={0 0 0},
}

\usepackage[todonotes={textsize=scriptsize},markup=underlined]{changes}
\definechangesauthor[name={R1}, color=black]{R1}
\definechangesauthor[name={R2}, color=black]{R2}
\definechangesauthor[name={R3}, color=black]{R3}
\setcommentmarkup{%
    \IfIsColored%
        {\colorlet{Changes@todocolor}{authorcolor}}%
        {\colorlet{Changes@todocolor}{black}}%
    \todo[color=Changes@todocolor!10, bordercolor=Changes@todocolor, linecolor=Changes@todocolor!70, nolist]{%
        \textbf{#3.} #1%
    }%
}

\newtheoremstyle{exampstyle}
  {.2em} 
  {.2em} 
  {\itshape} 
  {} 
  {\bfseries} 
  {.} 
  {.5em} 
  {} 

\ifCLASSINFOpdf
\else
\fi
\begin{document}
%
\title{Modeling and  Monitoring of Indoor Populations using Sparse Positioning Data (Extension)}

\author{Xiao Li,
        Huan Li,~\IEEEmembership{Member,~IEEE},
        Hua Lu,~\IEEEmembership{Senior~Member,~IEEE},
        and Christian S. Jensen,~\IEEEmembership{Fellow,~IEEE}
\IEEEcompsocitemizethanks{\IEEEcompsocthanksitem X.~Li and H.~Lu are with the Department
of People and Technology, Roskilde University, Denmark.
E-mail: \{xiaol, luhua\}@ruc.dk
\IEEEcompsocthanksitem H.~Li is with the College of Computer Science and Technology, Zhejiang University, China. E-mail: lihuan.cs@zju.edu.cn
\IEEEcompsocthanksitem C. S.~Jensen is with the Department of Computer Science, Aalborg University, Denmark.
E-mail: csj@cs.aau.dk}
}

\IEEEtitleabstractindextext{%
\begin{abstract}
In large venues like shopping malls and airports, knowledge on the indoor populations fuels applications such as business analytics, venue management, and safety control. 
In this work, we provide means of modeling populations in partitions of indoor space offline and of monitoring indoor populations continuously, by using indoor positioning data.  
However, the low-sampling rates of indoor positioning render the data temporally and spatially sparse, which in turn renders the offline capture of indoor populations challenging.
It is even more challenging to continuously monitor indoor populations,  as positioning data may be missing or not ready yet at the current moment.
To address these challenges, we first enable probabilistic modeling of populations in indoor space partitions as Normal distributions. Based on that, we propose two learning-based estimators for on-the-fly prediction of population distributions. Leveraging the prediction-based schemes, we provide a unified continuous query processing framework for a type of query that enables continuous monitoring of populated partitions.  
The framework encompasses caching and result validity mechanisms to reduce cost and maintain monitoring effectiveness. Extensive experiments on two real data sets show that the proposed estimators are able to outperform the state-of-the-art alternatives and that the query processing framework is effective and efficient.
\end{abstract}

}

\maketitle

\IEEEdisplaynontitleabstractindextext

%
\IEEEpeerreviewmaketitle

\section{Introduction}

According to several studies~\cite{jenkins1992activity, ott1988human}, people spend nearly 90\% of their lives in indoor spaces such as office buildings, shopping malls, metro stations, and airports. In many such indoor venues, keeping track of the populations of partitions\footnote{A partition is a unit of space like a room, a hallway, a metro platform, or an airport security area.} is of interest and importance for different purposes such as location-based services for end users~\cite{cheema2018indoor}, business analytics for shops~\cite{yaeli2014understanding},
and security and general venue management~\cite{li2018search}.
Knowing \textbf{indoor populations} has become even {more important in the post-COVID-19 era}, because it can contribute to enabling effective countermeasures, e.g., restricting the crowd size in the same room and enforcing social distancing~\cite{world2020covid}.

In this study, we provide means of modeling populations in indoor partitions in offline mode and of monitoring indoor populations continuously. 
Indoor populations are hard to capture due to the low quality of indoor positioning data, the counterpart of outdoor GPS data. 
As GPS does not work in indoor settings, indoor positioning systems are often based on short-range wireless technologies like Wi-Fi and Bluetooth~\cite{liu2007survey,li2023data}.
To the best of our knowledge, this paper is the first work of modeling populations in partitions of indoor space
offline and of monitoring indoor populations continuously, by using indoor positioning data.

{The indoor positioning systems are often characterized by a low sampling frequency, resulting in considerable uncertainties in the positioning data they produce. If we attempt to reconstruct the trajectory of an object (e.g., a shopper in a mall) from historical indoor positioning data, we may obtain a trajectory in which consecutively reported locations are far apart, temporally and (therefore) spatially. Such data sparsity is significant for applications in relatively small indoor spaces with compact partitions, a setting that is much different from that of outdoor GPS-based trajectories~}\cite{wang2019fast,tang2019joint}.
{Although studies}~\cite{zheng2012reducing,wu2016probabilistic} {on route recovery from sparse GPS data exist, they do not infer object locations at specific times, which is,  however, the key to knowing partition-level populations at a moment.
Moreover, unconstrained and road network-constrained movements in outdoor settings are quite different from indoor movements that are enabled or constrained by unique indoor entities like doors, walls, and partitions}~\cite{jensen2009graph,lu2012foundation}.
In general, the sparsity of indoor position data and the distinctive natures of movement in indoor space make it challenging to capture indoor populations.

First, it is non-trivial to know historical indoor populations. For example, given an indoor trajectory database, if we want to figure out the population of a partition (e.g., a shop in a mall) at a certain past time for business analytics, it is impossible to determine an accurate population via direct counting or statistics from the trajectory database. This is because inter-partition movements are not fully observed in the \emph{discrete, uncertain} indoor positioning data (to be detailed in Section~\ref{sec:modeling_population}). To contend with this, we conduct a theoretical analysis of inter-partition movements, {proving} that a partition's population at a certain time can be effectively captured as a Normal distribution. Moreover, we propose a concrete approach to compute the Normal population distributions of partitions, which involves a bound analysis of the time for an object to pass a door along a path as well as Monte Carlo sampling.

Next, it is even more challenging to monitor indoor populations \emph{online}. First, we cannot employ just mentioned population modeling approach to monitoring real-time populations because that approach requires follow-up observations that are unavailable during monitoring in real-time.
Second, due to the low sampling rate of indoor positioning, counting partition populations on the fly based on recently observed object locations will lead to low accuracy (cf.\ Section~\ref{ssec:query_processing}). %
Instead, we propose a prediction-based scheme to infer on-the-fly population distributions of partitions via learning from historical population distributions that are obtained by our proposed population modeling approach.

In particular, we design two neural network-based estimators to estimate  Normal distribution based populations in real-time. A single-way estimator works for each single partition by capturing the temporal dependencies of its historical populations, whereas a multi-way estimator works for all partitions together by further considering the spatial dependencies between connected partitions.
Both estimators employ a multi-task paradigm to predict the mean and variance of a Normal distribution jointly.

The resulting predicted population distributions enable us to find probabilistically populated partitions, i.e., those with a sufficiently high population.
Accordingly, we define a new type of query called \textbf{continuous monitoring of populated partitions} (CMPP). It continuously returns all populated partitions within a specified indoor distance of an object. 
Such a query finds application scenarios such as navigation and safety control.
For example, in a shopping mall or event venue, the {CMPP} can enable services that allow users to navigate to destinations via the fastest or safest route~\cite{liu2021towards} while taking into account information about crowds and congestion. 
As another example, in a large industrial facility or warehouse occupied by both human workers and equipment-carrying moving robots, the {CMPP} can be used to optimize the routes and workflows of the robots, improving workplace safety~\cite{inam2018safety}.

However, processing {CMPP} queries is challenging even when using the techniques above. In the prediction-based continuous monitoring setting, a population estimator continually needs feature sequences from recent data as input in order to make near-future population predictions. It is costly to generate feature sequences in real-time and to make frequent predictions. To contend with this, we propose a query processing framework that exploits the characteristics of the population estimators. In particular, we design a \emph{feature sequence caching} mechanism that maintains the previous feature sequences and a \emph{result validity} mechanism that reuses relevant and valid results from previous predictions. These two techniques reduce query processing costs substantially.

We evaluate our proposals on two real-world data sets. The results demonstrate that the learned population estimators are effective at predicting partition populations and the estimator-based continuous monitoring framework is effective and efficient at returning nearby populated partitions continuously.

In summary, we make the following main contributions:
\begin{itemize}[leftmargin=*]
    \item Based on theoretical analysis, we propose a probabilistic method that approximates an indoor partition's population at a past time {as a Normal distribution}.
    \item Based on the probabilistic method, we design two estimators to predict indoor partitions' future populations. 
    \item  Using the two estimators, we design a unified monitoring framework with validity and caching mechanisms for {continuous monitoring of populated partitions ({CMPP})}.
    \item We conduct extensive experiments on real data sets to evaluate our proposals.
\end{itemize}
The rest of this paper is organized as follows. Section~\ref{sec:prel} gives preliminaries and formulates problems. Section~\ref{sec:modeling_population} presents the probabilistic method for modeling indoor populations. Section~\ref{sec:estimators} and~\ref{sec:query_processing} 
focus on monitoring indoor populations. Specifically, Section~\ref{sec:estimators} proposes two indoor population estimators; Section~\ref{sec:query_processing} elaborates on the continuous monitoring based on the two estimators. Section~\ref{sec:exp} reports on extensive experiments. 
Section~\ref{sec:related} reviews related work. Section~\ref{sec:conclusion} concludes the paper and discusses future research directions.
\section{Preliminaries}
\label{sec:prel}

Table~\ref{tab:notations} lists notation used frequently in this paper.

\begin{table}[!htbp]
\footnotesize
\centering
\caption{Notation.}\label{tab:notations}
\begin{tabular}{cc}
\toprule
{Symbols} & {Meaning}\\
\midrule
$l$& an indoor location\\
$\phi$ & an indoor path\\
$\Phi$ & a set of possible paths\\
$v \in V$ & an indoor partition\\
$Pr(o \mid v, t)$ & probability that object $o$ is present at $v$ at time $t$\\
$Pr( v  \mid \phi, t)$ & probability that $\phi$ passes $v$ at time $t$\\
$P_{v,t}$ & population of $v$ at time $t$\\
\bottomrule
\end{tabular}
\end{table}

\subsection{Indoor Space and Positioning Data}
\label{ssec:space_and_data}

\begin{figure}[!htbp]
\centering
\includegraphics[width=0.9\columnwidth]{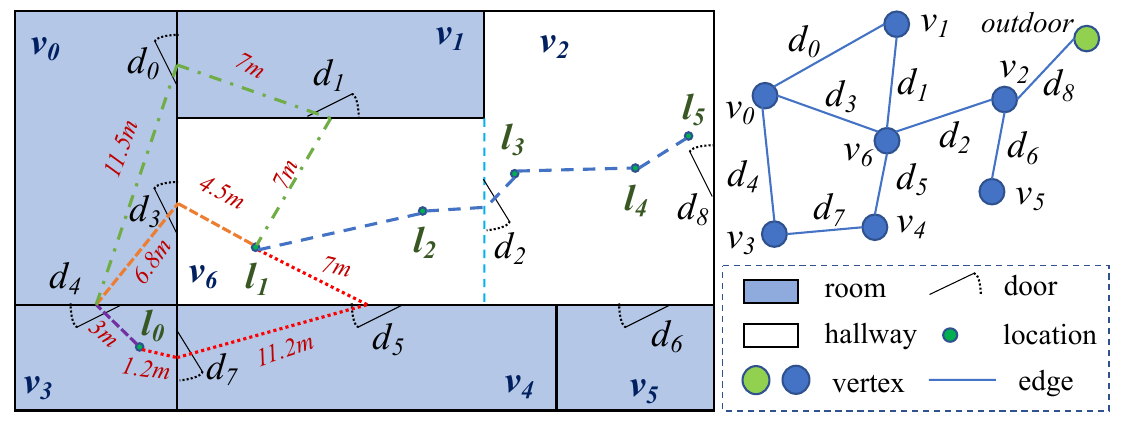}
\caption{Indoor floor plan and indoor graph.}
\label{fig:example}
\end{figure}

An indoor space is composed of partitions that are topological units. Two adjacent partitions are connected via a door.
Fig.~\ref{fig:example} illustrates an indoor space with 7 partitions \{$v_0, \ldots, v_6$\} and 9 doors \{$d_0, \ldots, d_8$\}.
Door $d_8$ connects partition $v_2$ to the outdoor space. The hallway is divided into two partitions $v_6$ and $v_2$ that are connected by door $d_2$.

To capture the connections among partitions and doors, we follow a previous study~\cite{lu2012foundation} and maintain the following mappings.
The mappings $P2D_{\sqsupset}(v_k)$ and  $P2D_{\sqsubset}(v_k)$ return the set of \emph{enterable} and \emph{leaveable} doors through which one can enter and leave the partition $v_k$, respectively.
Conversely, the mappings $D2P_{\sqsupset}(d_i)$ and  $D2P_{\sqsubset}(d_i)$ return the set of enterable and leaveable partitions of a door $d_i$, respectively. In Fig.~\ref{fig:example}, we have 
$P2D_{\sqsupset}(v_0) = \{d_0, d_3, d_4\}$,
$P2D_{\sqsubset}(v_1) = \{d_0, d_1\}$,
$D2P_{\sqsupset}(d_0) = \{v_0, v_1\}$, and
$D2P_{\sqsubset}(d_3) = \{v_0, v_6\}$.

Next, we use $v(l_i)$ to return the \emph{host partition} covering a location $l_i$, e.g., $v(l_0) = v_3$.
In addition to the mappings, the indoor topology is abstracted as a graph $G(V, E)$,  where each vertex in the set $V$ represents to a distinct indoor partition and each edge in the set $E$ corresponds to a door that connects two adjacent partitions (vertices).
An example indoor graph is shown in Fig.~\ref{fig:example} (top right).
An object can move freely inside a partition without changing the partition population. A change only occurs when an object enters another partition via a door. Therefore, we use an \emph{indoor path} to model an object's inter-partition movements from its source indoor location\footnote{An indoor location is generally a 2D point on a certain floor.} to its target indoor location.

\begin{definition}[Indoor Path] 
    An indoor path is represented as an interconnected door sequence between a source location $l_s$ and a target location $l_e$. Formally, $\phi = \langle l_s, d_1, \ldots, d_m, l_e \rangle$ where $v(l_s) \in D2P_{\sqsubset}(d_1)$, $v(l_e) \in D2P_{\sqsupset}(d_m)$, and $\forall d_i, d_{i+1} \in \phi$ ($1 \leq i < m$), $D2P_{\sqsubset}(d_i) \cap D2P_{\sqsupset}(d_{i+1}) \neq \varnothing$.
    The length of path $\phi$ is defined as:
    \begin{equation}
    \label{indoor_path}
     L(\phi) = |l_s,d_1|_E+\sum\nolimits_{i=1}^{m-1}|d_i,d_{i+1}|_E + |d_m,l_e|_E,
    \end{equation}
    where $|\cdot|_E$ denotes the Euclidean distance. 
\end{definition}
{We use the Euclidean distance for ease of understanding in the above definition. However, in cases where a partition contains obstacles, the concept of obstructed distance~\cite{DBLP:conf/edbt/ZhangPMZ04} can be applied locally to more accurately reflect the length of a path. 
It is noteworthy that an indoor path, as defined here, does not represent an actual trajectory taken by individuals. Instead, an indoor path serves as a concept for facilitating the analysis of population movements and distributions. In real-world scenarios, the actual trajectory between a source and a target location can vary significantly. Despite these variations, different real-world trajectories may be represented by the same indoor path in our abstract model.}

In our problem setting, an indoor positioning system reports positioning records of the form  $(o, l, t)$, meaning that a moving object $o$ is estimated to be at a location $l$ at a time $t$.
All historical positioning records are stored in a database.

Given an object, we obtain its \textbf{trajectory} as a time-ordered sequence of its reported locations recorded in the database, formally $\mathit{tr}$ = $\langle (l_1, t_1), \ldots, (l_n, t_n) \rangle$, with $t_1 < t_2 < \ldots < t_n$.
Partly due to the discrete nature of the reported locations in $\mathit{tr}$, it is necessary to project $\mathit{tr}$ into the corresponding indoor path to facilitate population counting.
However, this procedure is non-trivial. Multiple possibilities may exist between two consecutively reported locations in the presence of an indoor topology.
Also, as the specific time to pass an in-between door is unknown, we cannot simply count a partition's accurate population at a particular time.

\begin{example}\label{example:trajectories} 
Fig.~\ref{fig:example} shows an object $o$'s 6 sequentially reported locations, i.e., \{$l_0, \ldots, l_5$\}.
Three possible indoor paths exist for $o$ to reach $l_1$ from $l_0$: $\phi_1 = \langle l_0, d_4, d_0, d_1, l_1 \rangle$, $\phi_2 = \langle l_0, d_4, d_3, l_1 \rangle$, and $\phi_3 = \langle l_0, d_7, d_5, l_1 \rangle$. As no extra evidence is given, $v_0$, $v_1$, and $v_4$ are all possible partitions that $o$ has passed.
Although $o$ went through $d_2$ during the time interval $(t_2, t_3)$, it is uncertain whether $o$ should be counted in $v_2$'s population at an arbitrary timestamp $t' \in (t_2, t_3)$.
\end{example}

To deal with the uncertainty in infrequently sampled trajectories, we employ a probabilistic method (see Section~\ref{sec:modeling_population}) to model inter-partition movements and populations.
As a result, we prove that a partition $v$'s population at a particular timestamp $t$, denoted as a random variable $P_{v,t}$, can be approximated by a Normal distribution.

\subsection{Problems {and Proposed Framework}}
\label{ssec:problem}

By applying Normal distributions to partition populations, we define \emph{populated partition} as follows.

\begin{definition}[Populated Partition]
    Given a population threshold $\theta \in \mathbb{Z}^{+}$ and a confidence threshold $\eta \in (0,1)$, a partition $v$ is a \emph{populated partition} at time $t$ if the \emph{probability mass function} (PMF) $f$ of its population $P_{v,t}$ satisfies:
    \begin{equation}
    \label{equation:pmf_populated}
    f(P_{v,t} > \theta) \geq \eta,
    \end{equation}
    i.e., $P_{v,t}$ exceeds $\theta$ with probability at least $\eta$. 
\end{definition}
Both $\theta$ and $\eta$ are user-defined parameters to decide if an indoor partition is populated or not. The setting of their values depends on users' demand in the concrete application scenarios.

We proceed to define our research problems as follows.
\begin{problem}[Population Modeling]\label{problem_1}
    Given a trajectory database $\mathit{TR} = \{\mathit{tr}_i = \langle (l_1, t_1), \ldots, (l_n, t_n) \rangle \}$ and a set $V^{*} \subseteq V$ of  concerned partitions, the \emph{population modeling} aims to decide the distribution of the population $P_{v,t}$ of each partition $v \in V^{*}$ at a specified historical timestamp $t$ within the scope of $\mathit{TR}$.

\end{problem}

\begin{problem}[Continuous Monitoring of Populated Partitions, {CMPP}]\label{problem_2} 
    Given a distance range $r$, a population threshold $\theta$, a confidence threshold $\eta$, an end time $t_\text{end}$, and a query time interval $\Delta t$, the \emph{continuous monitoring of populated partitions} query $\texttt{CMPP}(r, \theta, \eta, t_\text{end}, \Delta t)$ returns a list of populated partitions satisfying the thresholds $\theta$ and $\eta$ within the range $r$ of the current query location when
    \setlength{\parskip}{0pt}
    \begin{enumerate}[topsep=0pt, partopsep=-\parskip, itemsep=0pt, parsep=0pt]
    \item the current time is no older than $t_\text{end}$, 
    \item the query location is changing, and
    \item the time since the most recent result update exceeds $\Delta t$.
    \end{enumerate}
\end{problem}

\begin{figure}[!htbp]
\centering
\includegraphics[width=0.92\columnwidth]{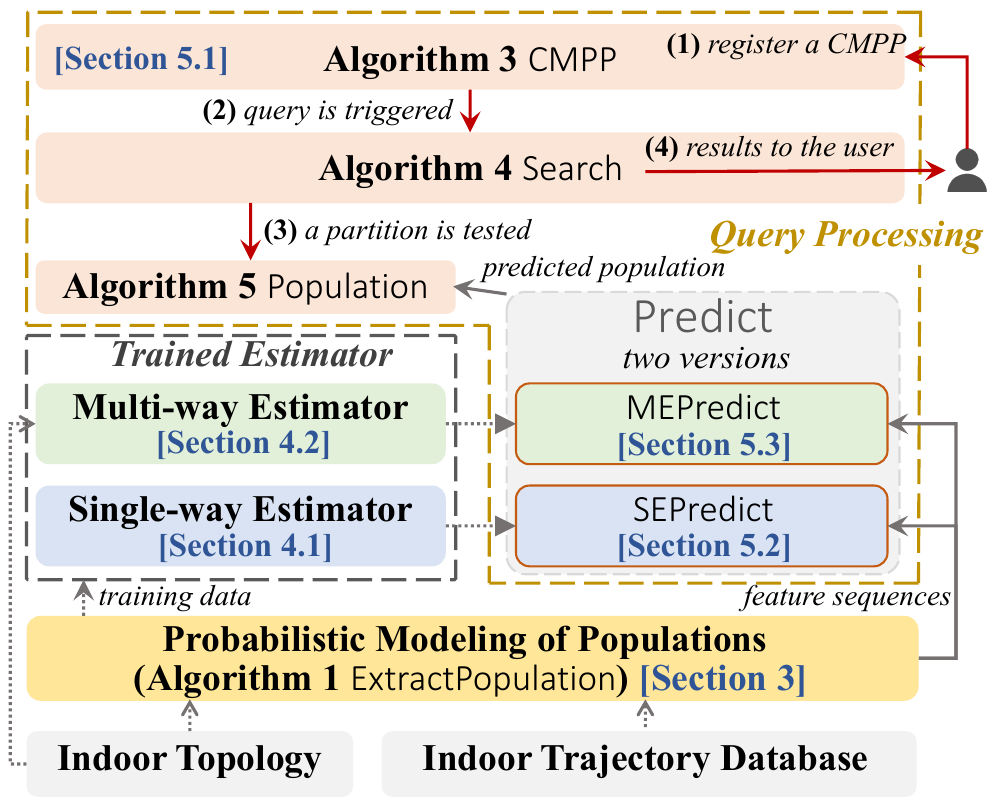}
\caption{Processing framework overview.}
\label{fig:framework}
\end{figure}

To solve the above problems, we propose a processing framework as shown in Fig.~\ref{fig:framework}. It illustrates the process from probabilistic indoor population modeling to estimator-based {CMPP} processing, which leverages indoor topology and an indoor trajectory database.

At the bottom of the framework, the indoor topology and trajectory database are needed to generate the Normal distribution based partition-level populations at historical timestamps. The generation of such probabilistic populations is implemented as an algorithm \textsc{ExtractPopulation} (Algorithm~\ref{alg:transform}),  to be detailed in Section~\ref{sec:modeling_population}.

We design two population estimators, both trained by using the historical population information extracted by Algorithm~\ref{alg:transform}.
The single-way estimator (Section~\ref{ssec:local_estimator}) predicts the population of a single partition. The multi-way estimator (Section~\ref{ssec:global_estimator}) utilizes the indoor topology to predict for multiple partitions simultaneously.

On top of the learned estimators, we design a uniform framework (Section~\ref{ssec:framework}) for continuous monitoring.
When a user registers a query, the main function \textsc{CMPP} (Algorithm~\ref{alg:CMPP}) monitors the current situation and calls a process \textsc{Search} (Algorithm~\ref{alg:search}) to find nearby populated partitions. 
Next, \textsc{Search} calls a process \textsc{Population} (Algorithm~\ref{alg:test}) to determine whether a reachable partition is populated. In \textsc{Population}, a candidate partition's population is fetched from either making a 
 prediction or the result just predicted by one of the estimators. Corresponding to the two estimators, two versions of \textsc{Predict} are designed, namely \textsc{SEPredict} (Section~\ref{ssec:se_predict}) and \textsc{MEPredict} (Section~\ref{ssec:me_predict}). 
For online prediction, Algorithm~\ref{alg:transform} is also evoked to generate feature sequences.
When a list of populated partitions is found by \textsc{Search}, the list is returned to the query user.

\section{Probabilistic Modeling of Populations}
\label{sec:modeling_population}

\subsection{{Populations Derived as Normal Distribution}}
\label{ssec:gaps_data}
As described in Section~\ref{ssec:space_and_data}, the indoor positioning data is sparse, i.e., two consecutive reported locations may be far apart from each other such that a number of paths are possible between them (see Example~\ref{example:trajectories}). In this section, we fill in such gaps and attach meaningful probabilities to such possible intermediate paths.
We represent the \textbf{possible cross-door movement} between two consecutive locations $l_A$ and $l_B$ in a trajectory as a random variable path $\phi = \langle l_A, d_1, \ldots, d_m, l_B \rangle$.
The variable $\phi$ simply follows a categorical distribution $\mathcal{D}$ with a probability mass function (PMF) $f$ as:
\begin{equation}\label{equation:phi}
    f(\phi = \phi_{j}) = \mathit{Pr}(\phi_{j}), \,\,j=1,\ldots,\mathtt{J},
\end{equation}
where $\mathtt{J}$ is the number of possible paths, $\mathit{Pr}(\phi_{j})$ is the occurrence probability of the $j$-th possible path $\phi_{j}$, and $\sum_{j=1}^{\mathtt{J}} \mathit{Pr}(\phi_{j}) = 1$ holds.

At a particular time $t$ between $l_A$'s observed time $t_A$ and $l_B$'s observed time $t_B$, an object $o$'s \textbf{presence} at a partition $v$ is captured as a probability: 
\begin{equation}\label{equation:presence}
\mathit{Pr}(o \mid v, t) = \sum\nolimits_{j=1}^\mathtt{J} \mathit{Pr}(\phi_{j}) \cdot \mathit{Pr}(v \mid \phi_{j}, t),
\end{equation}
where $\mathit{Pr}(v \mid \phi_{j}, t) \in [0,1]$ models the probability that path $\phi_{j}$ passes partition $v$ at time $t$.\footnote{For any partition $v$ that $\phi_{j}$ has not passed, $\mathit{Pr}(\phi_{j} \mid v, t)$ is 0.}
Due to the randomness of movement on $\phi_{j}$, $\mathit{Pr}(v \mid \phi_{j}, t)$ is non-deterministic.
However, we estimate it based on the Monte Carlo sampling over the possible time interval within which the object passes a door along $\phi_{j}$ (to be detailed in Section~\ref{ssec:transformation}).

Simply, object $o$'s presence at $v$ at time $t$ is a Bernoulli variable $\Omega_{o, v,t} \in \{0, 1\}$ with success probability $\mathit{Pr}(o \mid v, t)$. The success probability is \emph{non-identical} across all objects.

Furthermore, we use $P_{v,t} = \sum_{o\in O} \Omega_{o, v,t}$ to denote the \textbf{population of a partition} $v$ at a time $t$. The value of $P_{v,t}$ is an integer $p$ between 0 and $|O|$ where $O$ is the full set of moving objects in the space at time $t$.
In other words,  $P_{v,t}$ has ($|O|$ + 1) possible categories, i.e., $p^{(0)}, \ldots, p^{(|O|)}$.
As a sum of non-identically distributed Bernoulli variables, $P_{v,t}$ follows the \emph{Poisson Binomial distribution}~\cite{wang1993number,fernandez2010closed} with the following PMF:  
\begin{equation}\label{equation:X}
\begin{aligned}
& f(P_{v,t} = p) = \\
& ~~\sum\limits_{C_p \in \mathbb{C}(p)} \prod\limits_{o_i \in C_p} \mathit{Pr}(o_i \mid v, t) \prod\limits_{o_j \in O \setminus C_p} \Big(1-\mathit{Pr}(o_j \mid v, t)\Big),
\end{aligned}
\end{equation}
where $\mathbb{C}(p)$ covers all possible combinations of the $p$ objects that are present in $v$ at time $t$, satisfying $|\mathbb{C}(p)|=\binom{|O|}{p}$, and
$C_p \in \mathbb{C}(p)$ refers to one possible combination.
According to the central limit theorem~\cite{hogg1977probability}, a Poisson Binomial distribution can be approximated as a {Normal distribution}.\footnote{A detailed theoretical analysis is available elsewhere~\cite{neammanee2005refinement}.}
As a result, we have:
\begin{equation}\label{equation:gaussian_approx}
\begin{aligned}
P_{v,t} & \sim \mathcal{N}(\mu, \sigma^2), \\
\mu & = \sum\nolimits_{o \in O} \mathit{Pr}(o \mid v, t), \\
\sigma^2 & = \sum\nolimits_{o \in O} \mathit{Pr}(o \mid v, t) \cdot (1 - \mathit{Pr}(o \mid v, t)).
\end{aligned}
\end{equation}
{To sum up, the partition population at a time $t$ is described by a mean $\mu$ and a variance $\sigma^2$.} The approximation accuracy depends mostly on the number of trials. In our context, the number of trials corresponds to that of the number of objects involved, $|O|$. Usually, the more objects are involved, the higher the accuracy. As a rule of thumb, a satisfactory approximation effect is obtained when $|O| \geq 30$~\cite{hogg1977probability}. This is not difficult to achieve in our context, where a large indoor space can easily accommodate many more objects.

In order to estimate ($\mu$, $\sigma^2$), we must generate $\mathit{Pr}(\phi_{j})$, i.e., $\phi_{j}$'s occurrence probability, and $\mathit{Pr}(v \mid \phi_{j}, t)$, i.e., the probability to pass partition $v$ via path $\phi_{j}$ at time $t$ (see Equation~\ref{equation:presence}).
We proceed to present a holistic approach to generating these probabilities for deriving Normal distribution-based populations.

\subsection{Generation of Probabilistic Populations}

\label{ssec:transformation}
By accessing the indoor trajectories, Algorithm~\ref{alg:transform} extracts the mean and variance of the population of each partition in a given set $V^{*}$ at a historical timestamp $t$. The algorithm also utilizes the maximum speed~$\mathcal{S}_{max}$ of all moving objects in the indoor space~\cite{willen2013walking}, which can be obtained via statistics of the historical data.
Initially, the algorithm retrieves each pair of consecutive reported locations from the trajectories, where each pair's corresponding time interval covers time $t$ (line~1).
Note that each object (trajectory) can have at most one such pair $(l_A, l_B)$.  Next,
for each pair $(l_A, l_B)$, a set $\Phi$ of possible indoor paths between them are found. 
Among them, we exclude any path $\phi$ satisfying $L{(\phi)}> \mathcal{S}_{max} \cdot (t_B - t_A)$, since its length exceeds the maximum walking distance by the object during the period (line 3). Next, each path $\phi_j \in \Phi$ gets a possibility $\mathit{Pr}(\phi_j)$ inversely proportional to $\phi_j$'s length (lines 4--5). Thus, a shorter path has a higher probability, in line with common sense.\footnote{Nevertheless, other factors like user preferences can be incorporated through a preference weight~\cite{yawalkar2019route} before normalization.}

\begin{algorithm}[!htbp]
\footnotesize
\caption{\footnotesize \textsc{ExtractPopulation}~(Partition set $V^{*}$, timestamp $t$, maximum speed $\mathcal{S}_{max}$)} \label{alg:transform}
\begin{algorithmic}[1]
\State retrieve from the trajectories all pairs of consecutive reported locations $(l_A, l_B)$ such that $t \in [t_A, t_B]$
\For{each retrieved pair $(l_A, l_B)$} \Comment{corresponds to an object $o$}
    \State path set $\Phi \gets$  $\{ \phi \text{~from~} l_A \text{~to~} l_B \mid L{(\phi)} \leq \mathcal{S}_{max} \cdot (t_B - t_A)$ \} 
    \For{each path $\phi \in \Phi$}
        \State probability $\mathit{Pr}(\phi) \gets \frac{1/L(\phi)}{\sum_{\Phi}1/L(\phi_i)}$
       
        \State  $v.\text{count}$ $\gets$ 0 for each {$v \in V^{*}$}
      
        \For{${\mathtt{k}=1}$ to $\mathtt{K}$}
        \Comment{to sample $\mathtt{K}$ times}
            \State $v' \gets$ \Call{FindPartition}{$l_A$, $l_B$,  $t_A$, $t_B$, $S_{max}$, $\phi$, $t$}
            \State $v'.\text{count} \gets v'.\text{count} + 1$
        \EndFor
        
        \For{each partition $v \in V^{*}$}
            \State $\mathit{Pr}(v \mid \phi, t) \gets v.\text{count}/\mathtt{K}$
        \EndFor
    \EndFor
    \For{each partition $v \in V^{*}$}
        \State compute $\mathit{Pr}(o \mid v, t)$ according to Equation~\ref{equation:presence}
    \EndFor
\EndFor
\For{each partition $v \in V^{*}$}
    \State compute $\mu_{v,t}$ and $\sigma_{v,t}^2$ according to Equation~\ref{equation:gaussian_approx}
\EndFor
\State \Return $(\mu_{v,t}, \sigma_{v,t}^2)$ for each $v \in V^{*}$
\end{algorithmic}
\end{algorithm}

Recall that the randomness of movement on a path $\phi_j$ complicates the determination of the time at which a moving object passes a door. Consequently, it is also difficult to determine the probability $\mathit{Pr}(v \mid \phi_j,t)$ in Equation~\ref{equation:presence}.
Nevertheless, we are able to calculate the bounds of the time to pass a door on $\phi_j$, based on which Monte Carlo sampling is used to find a possible partition where the object is at time $t$. All this is done in \textsc{FindPartition} (line~8, to be detailed in Algorithm~\ref{alg:findpartition}).
We conduct the sampling $\mathtt{K}$ times and count the frequency of presence of each partition (lines 7--9). Consequently, the probability $\mathit{Pr}(v \mid \phi,t)$ for each partition is calculated as the frequency of presence of each partition divided by $\mathtt{K}$ (lines 10--11).
At each iteration on a pair $(l_A, l_B)$, the probability $\mathit{Pr}(o \mid v, t)$ is calculated according to Equation~\ref{equation:presence} (lines~12--13).
After the iterations, the algorithm estimates the mean $\mu_{v,t}$ and variance $\sigma_{v,t}^2$ for each partition $v$ based on $\mathit{Pr}(o \mid v, t)$ over all its relevant objects (lines~14--15).
The estimated results are returned for all partitions in $V^{*}$ (line~16).

Algorithm~\ref{alg:findpartition} \textsc{FindPartition} finds a possible partition for the object at time $t$ on a path $\phi$. The basis is to infer the time to pass each door on that path. We initialize the starting location $l_s$ as $l_A$ and its time $t_s$ as $t_A$ (line 1). In lines~2--6, we iterate through each door on $\phi$. In each iteration, we derive the lower and upper bounds of the time to pass a door $d_k$ (lines~3--4). Specifically, the \textbf{lower bound} is derived by assuming the object takes the shortest distance from the starting location $l_s$ to the current door $d_k$, denoted as $L(l_s \leadsto d_k)$ in line~3, at the maximum speed. In contrast, the \textbf{upper bound} is obtained by assuming the object takes the shortest distance from the current door $d_k$ to the end of this path $l_B$, i.e., $L(d_k \leadsto l_B)$ in line~4, such that the most time is used before $d_k$. Subsequently, following the principle of Monte Carlo method, we randomly sample a time within the derived bounds as the time to pass that door $d_k$ (line 5). Furthermore, the current door and its newly generated time are taken as the new starting location and its time, respectively, for the next iteration (line 6). The iterations find the time to pass each door on path $\phi$, which enables us to know the two consecutive doors between which the object is at the time $t$ (line~7). Finally, the passed partition between the two doors is found and returned (lines~8--9).

\begin{algorithm}[!htbp]
\footnotesize
\caption{\footnotesize \textsc{FindPartition}~($l_A$, $l_B$,  $t_A$, $t_B$, $S_{max}$, $\phi$, $t$)} \label{alg:findpartition}
\begin{algorithmic}[1]
    \State  $t_s \gets$ $t_A$; $l_s \gets$ $l_A$
    \For{each door $d_k$ on $\phi$}
    \Comment{iteratively obtain $t_k$ w.r.t. $d_k$}
        \State $t_k^\mathit{LB} \gets t_s + \frac{L(l_s \leadsto d_k)}{S_{max}}$
        \Comment{lower bound of $t_k$}
        \State  $t_k^\mathit{UB} \gets t_B - \frac{L(d_k \leadsto l_B)}{S_{max}}$
    \Comment{upper bound of $t_k$}
        \State pass time $t_k \gets$ a random time within $[t_k^\mathit{LB}, t_k^\mathit{UB}]$
        \State $t_s \gets t_k$; $l_s \gets d_k$
        \Comment{$d_k, t_k$ as the new starting location and time}
    \EndFor
    \State find $d_i$ and $d_{i-1}$ such that $t \in [t_{i-1}, t_i]$
    \State $v' \gets D2P_{\sqsubset}(d_{i-1}) \cap D2P_{\sqsupset}(d_{i})$
\State \Return $v'$
\end{algorithmic}
\end{algorithm}

Next, we give an example to illustrate how to leverage Algorithm~\ref{alg:transform}, which calls Algorithm~\ref{alg:findpartition} on line~8, to generate Normal-based partition populations.

\begin{example}\label{example:sample_based_trajectories} 
Continuing with Example~\ref{example:trajectories} and Fig.~\ref{fig:example}, for the three paths between $l_0$ and $l_1$ (cf.\ Example~\ref{example:trajectories}), their lengths are L($\phi_1$) = 28.5 m, L($\phi_2$) = 14.3 m, and L($\phi_3$) = 19.4 m, respectively. 
Assume $t_0 = 0$ s, $t_1= 15$ s, $t = 5$ s, and $S_{max}=1.53$ m/s. 
We first exclude $\phi_1$ as $1.53*(15-0)$ = 22.95 m~$<$~L($\phi_1$) = 28.5 m.
Next, according to line~5 in Algorithm~\ref{alg:transform}, we obtain $\mathit{Pr}(\phi_2)=\frac{14.3^{-1}}{14.3^{-1} + 19.4^{-1}} = 0.576$, 
and likewise $\mathit{Pr}(\phi_3)=0.424$.
We take $\phi_2 = \langle l_0, d_4, d_3, l_1 \rangle$ as an example to show which partition the object would be at time $t$. 
For the first door $d_4$, the lower bound of the time for the object to pass is {$0 + \frac{3}{1.53} =  1.96$} s ($L(l_0 \leadsto d_4)$ = 3 m) and the upper bound is {$15 - \frac{11.3}{1.53} = 5.65$} s ($L(d_4 \leadsto l_1)$ = 6.8 + 4.5 = 11.3 m).

We then sample a random time within $[1.96, 5.65]$ as the time to pass $d_4$, say $2.5$ s. 
Based on that, we further derive the bounds for the next door $d_3$ as $[2.5+\frac{6.8}{1.53}, 15-\frac{4.5}{1.53}]$ = $[6.94, 12.06]$ ($L(d_4 \leadsto d_3)$ = 6.8 m and $L(d_3 \leadsto l_1)$ = 4.5 m). 
Likewise, we sample the time to pass $d_3$ within $[6.94, 12.06]$, say $7.5$ s.
In this sampling, the object passes $d_3$ at 2.5 s and $d_4$ at 7.5 s, which indicates it is located in partition $v_0$ at the time $t$ = 5 s (see Fig.~\ref{fig:example}).
Suppose the sampling is performed for $\mathtt{K}=100$ times and the counts of possible located partitions $v_3$, $v_0$, and $v_6$ at $t$ are 10, 40, and 50, respectively.
We have $\mathit{Pr}(v_3 \mid \phi_{2}, t) =0.1$, $\mathit{Pr}(v_0 \mid \phi_{2}, t) =0.4$, and $\mathit{Pr}(v_6 \mid \phi_{2}, t) =0.5$. For the partitions that $\phi_2$ never passes, their probabilities are set to 0.
Following the same way as for $\phi_{2}$, suppose we obtain $\mathit{Pr}(v_3 \mid \phi_{3}, t) =0.05$, $\mathit{Pr}(v_4 \mid \phi_{3}, t) =0.65$, and $\mathit{Pr}(v_6 \mid \phi_{3}, t) =0.3$ for $\phi_{3}$. 

According to Equation~\ref{equation:presence}, $\mathit{Pr}(o \mid v_3, t) = 0.576 \cdot 0.1 + 0.424 \cdot 0.05 = 0.0788$. Likewise, $\mathit{Pr}(o \mid v_0, t) = 0.2304$, $\mathit{Pr}(o \mid v_6, t) = 0.4152$, and $\mathit{Pr}(o \mid v_4, t) = 0.2756$. 
Assume there is only one object involved. 
Finally, based on Equation~\ref{equation:gaussian_approx}, the populations of relevant partitions are approximated as $P_{v_3,t} \sim \mathcal{N}(0.0788, 0.0726)$, $P_{v_0,t} \sim \mathcal{N}(0.2304, 0.0.1773)$, $P_{v_4,t} \sim \mathcal{N}(0.2756, 0.1996)$, and $P_{v_6,t} \sim \mathcal{N}(0.4152, 0.2428)$.
\end{example}
\section{Indoor Population Estimators}
\label{sec:estimators}

{We propose two estimators for predicting on-the-fly partition populations (the mean $\mu_{v,t}$ and the standard deviation $\sigma_{v,t}$).}
A \textbf{single-way estimator} predicts the population for only one partition, while a \textbf{multi-way estimator} can predict the populations for all partitions in one pass.

\subsection{Single-way Estimator}
\label{ssec:local_estimator}

\begin{figure}[!htbp]
\centering
\includegraphics[width=\columnwidth]{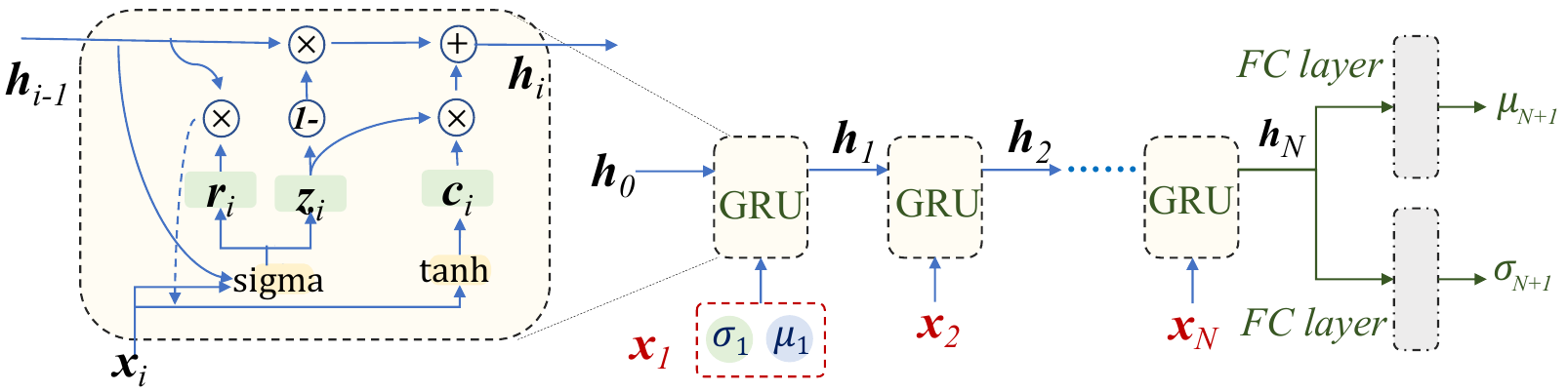}
\caption{Architecture of single-way estimator (SE).}
\label{fig:arch_se}
\end{figure}

Given a partition $v$, a single-way estimator (SE) takes a sequence of $v$'s $N$ most recent historical populations as the input (i.e, feature sequence), and outputs the mean $\mu_{v,t}$ and standard deviation $\sigma_{v,t}$ of the population at the current timestamp $t$. 
In order to exploit the temporal dependencies in historical populations,
{SE extends a Gated Recurrent Unit (GRU) network~\cite{chung2014empirical} that is effective at capturing temporal dependencies of data~\cite{zhao2019t}.}
However, SE uses two parallel fully connected (FC) layers to output $\mu_{v,t}$ and $\sigma_{v,t}$ respectively, as depicted in Fig.~\ref{fig:arch_se}.
The lower layers in GRUs are shared in SE for predicting the mean and variance, and the whole SE is trained based on a multi-task learning paradigm. SE's main components are detailed as follows.

\smallskip
\noindent\textbf{Input.} Let $\delta$ be the time window size between two consecutive GRUs.
The $i$-th ($1 \leq i \leq N$) input population is captured as:
\begin{equation}\label{equation:gru_input}
\mathbf{x}_{i} = [\mu_{v,t-(N-i+1)\delta}, \sigma_{v,t-(N-i+1)\delta}]
\end{equation}

\smallskip
\noindent\textbf{Internal.}
The internal is $N$ stacked GRUs that altogether capture the temporal dependencies of the populations over the time. Based on the following formulas, the $i$-th ($1 \leq i \leq N$) unit takes $\mathbf{x}_{i}$ and latent vectors from the previous unit and outputs the latent vectors to be fed into the next unit.
\begin{equation}\label{equation:gru}
\begin{aligned}
\mathbf{z}_i = & \; \text{sigmoid}( \mathbf{W}_z \mathbf{x}_i + \mathbf{U}_z \mathbf{h}_{i-1} + \mathbf{b}_z)\\
\mathbf{r}_i = & \; \text{sigmoid}(\mathbf{W}_r \mathbf{x}_i + \mathbf{U}_r \mathbf{h}_{i-1} + \mathbf{b}_r)\\
\mathbf{c}_i = & \; \tanh(\mathbf{W}_c \mathbf{x}_i + \mathbf{U}_h (\mathbf{r}_i \odot \mathbf{h}_{i-1}) + \mathbf{b}_c) \\
\mathbf{h}_i = & \; \mathbf{z}_i \odot \mathbf{c}_{i} + (1-\mathbf{z}_i)\odot \mathbf{h}_{i-1},\\
\end{aligned}
\end{equation}
where $\mathbf{z}_i$ is the update gate vector capturing the long-term memory, $\mathbf{r}_i$ is the reset gate vector capturing the short-term memory, $\mathbf{c}_i$ is the candidate activation vector, $\mathbf{h}_i$ is the  output vector, $\odot$ denotes Hadamard product operator, and $\text{sigmoid}(\cdot)$ and $\tanh(\cdot)$ are activation functions.
All weight matrices $\mathbf{W}_{*}$, $\mathbf{U}_{*}$ and bias vectors $b_{*}$ in Equation~\ref{equation:gru} are learnable parameters. 
Through the $N$ GRUs, the final output $\mathbf{h}_{N}$ at the $N$-th unit can encode long- and short-term temporal dependencies between the input sequential populations.

\smallskip
\noindent\textbf{Output.}
{Given the output vector $\mathbf{h}_{N}$ from the final GRU, the mean and standard deviation are predicted through two FC layers:}
\begin{equation*}
\hat{\mu}_{v,t} = \mathbf{w}_{\mu} \mathbf{h}_N + b_{\mu}; \;\:\;\;
\hat{\sigma}_{v,t} = \mathbf{w}_{\sigma} \mathbf{h}_N + b_{\sigma},
\end{equation*}
{where $\hat{\mu}_{v,t}$ and $\hat{\sigma}_{v,t}$ denote the predicted mean and standard deviation.}

\noindent\textbf{Loss Function.} Since the model's output denotes a Normal distribution, the loss is defined on the distance between the predicted Normal distribution (i.e., $\widehat{P}_{v,t} \sim \mathcal{N}(\hat{\mu}_{v,t}, \hat{\sigma}_{v,t}^2)$) and the ground-truth Normal distribution (i.e., $P_{v,t} \sim \mathcal{N}(\mu_{v,t}, \sigma_{v,t}^2)$). First, to measure the distance between the two Normal distributions, we introduce a typical metric called 2-Wasserstein distance~\cite{hettige2020robust} as follows:

\begin{equation}
    W_2(\widehat{P}_{v,t}, P_{v,t}) = (||\mu_{v,t} - \hat{\mu}_{v,t}||_2^2 + ||\sigma_{v,t} - \hat{\sigma}_{v,t}||_{F}^2)^{\frac{1}{2}},
\end{equation}
where $\mu_{v,t}$ and $\sigma_{v,t}$ are the ground-truth mean and standard deviation of the population distribution respectively, and $F$ represents the Frobenius norm. Since the population of a partition is modeled as a univariate Normal (i.e., Gaussian) distribution instead of a multivariate Normal distribution,  $\mu_{v,t}$, $\hat{\mu}_{v,t}$, $\sigma_{v,t}$, and $\hat{\sigma}_{v,t}$ are all scalar values. This makes $||\sigma_{v,t} - \hat{\sigma}_{v,t}||_{F}^2$ equal $||\sigma_{v,t} - \hat{\sigma}_{v,t}||_{2}^2$. 
Moreover, to render the loss function differentiable, the squared Wasserstein distance is used as the loss function as follows:
\begin{equation}\label{equation:wa_loss}
\begin{aligned}
\mathcal{L}_\text{SE} =  W_2(\widehat{P}_{v,t}, P_{v,t})^2 = ||\mu_{v,t} - \hat{\mu}_{v,t}||_2^2 + ||\sigma_{v,t} - \hat{\sigma}_{v,t}||_{2}^2
\end{aligned}
\end{equation}
Notably, the loss turns out to be the sum of Mean Square Errors (MSEs) for both mean and standard deviation predictions. We discuss the choice of this loss function and possible alternatives in Appendix~\ref{ssec:loss_func}.

SE is trained based on historical populations in a single partition, which does not exploit population information from adjacent partitions.
However, a partition's current population is related to its adjacent partitions' populations at previous times because objects in one partition can move from or to its adjacent partitions. 
To this end, we proceed to present an improved estimator.

\subsection{Multi-way Estimator}
\label{ssec:global_estimator}

\begin{figure}[!htbp]
\centering
\includegraphics[width=\columnwidth]{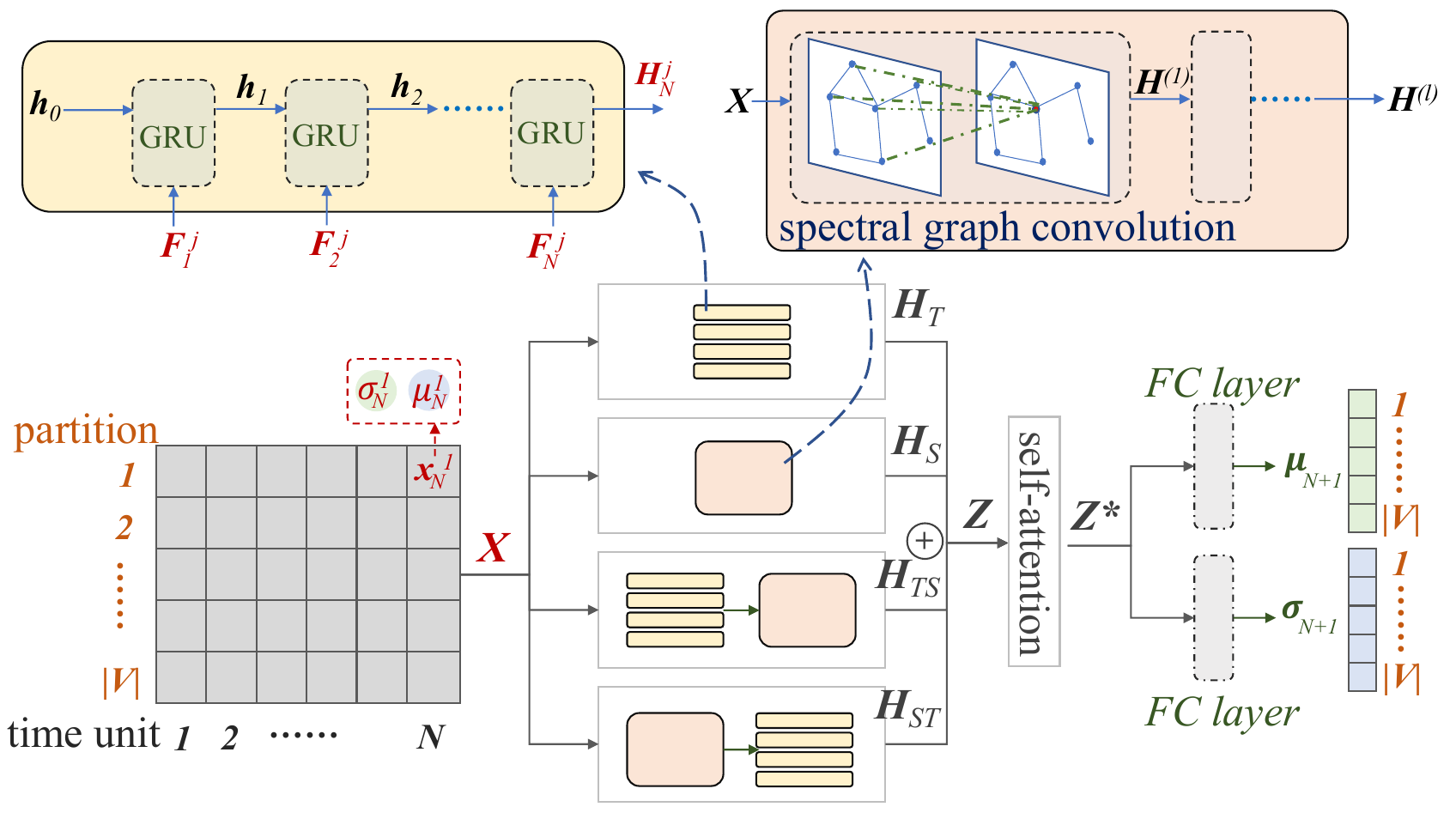}
\caption{Architecture of multi-way estimator (ME).}
\label{fig:me}
\end{figure}

A multi-way estimator (ME) takes all partitions' population information at the $N$ recent historical timestamps as the input, and outputs their current populations in one pass.
The ME architecture is depicted in Fig.~\ref{fig:me}.
ME comprehensively captures the temporal dependencies among sequential populations and spatial dependencies among adjacent partitions.

\smallskip
\noindent\textbf{Input.}
Let $V$ be the set of all partitions.
The populations at the $i$-th ($1 \leq i \leq N$) timestamp is represented as
\begin{equation*}
\mathbf{X}_{i} = [\mathbf{x}^1_i, \ldots, \mathbf{x}^{|V|}_i]^{\mathsf{T}},
\end{equation*}
where $\mathbf{x}^{j}_i$ ($1 \leq j \leq |V|$) is the $j$-th partition's population information (i.e., mean and variance) and carries the same meaning as that in Equation~\ref{equation:gru_input}.
The full input is represented as $\mathbf{X} = [\mathbf{X}_1, \ldots, \mathbf{X}_N]^{\mathsf{T}}$.
The size of $\mathbf{X}$ is $2 \times |V| \times N$.
We denote partition $v_j$'s population information at all $N$ timestamps as $\mathbf{X}[:,j,:] \in \mathbb{R}^{2 \times N}$ or simply $\mathbf{X}^j$, and the $i$-th historical timestamp's population information at all partitions as $\mathbf{X}[:,:,i] \in \mathbb{R}^{2 \times |V|}$ or simply $\mathbf{X}_i$.

\smallskip
\noindent\textbf{Internal.}
ME's internal consists of four parallel units for the temporal, spatial, temporal-spatial, and spatial-temporal dependencies, respectively, and a self-attention unit to integrate the four kinds of dependencies for predicting the populations.

The \textit{Temporal Unit} is generalized as a function $\texttt{T}: [\mathbf{F}_1, \ldots, \mathbf{F}_N]$ $\mapsto \mathbf{H}$ that maps the input tensor $\mathbf{F}_i \in \mathbb{R}^{|V| \times K'}$ at $N$ sequential timestamps to a latent tensor $\mathbf{H} \in \mathbb{R}^{|V| \times K}$.
We instantiate $\texttt{T}$ as a $|V|$-way vanilla GRU network with $N$ units. Each way processes on the input sequence $[\mathbf{F}^j_1, \ldots, \mathbf{F}^j_N ]$ of the $j$-th partition where $\mathbf{F}^j_i$ is the $j$-th row of $\mathbf{F}_i$, and outputs $\mathbf{H}^j$ as the $j$-th row of $\mathbf{H}$.

The \textit{Spatial Unit} is generalized as a function $\texttt{S}: [\mathbf{F}^1, \ldots, \mathbf{F}^{|V|}]^{\mathsf{T}}$ $\mapsto \mathbf{H}$ that maps the feature tensors at all $|V|$ partitions to a latent tensor $\mathbf{H} \in \mathbb{R}^{|V| \times K}$.
We instantiate $\texttt{S}$ as a $\ell$-layer graph convolution network (GCN) and take the input organized as $\mathbf{X} = [\mathbf{X}^1, \ldots, \mathbf{X}^{|V|}]$.

The GCN transforms the input $\mathbf{X}$ to the hidden vector based on the spectral graph convolution~\cite{kipf2016semi} formulated as follows.
\begin{equation*}
\begin{aligned}
& \mathbf{H}^{(\ell)} = \text{ReLu}(\tilde{\mathbf{D}}^{-\frac{1}{2}}\tilde{\mathbf{A}}\tilde{\mathbf{D}}^{-\frac{1}{2}} \mathbf{H}^{(\ell-1)} \mathbf{W}^{(\ell-1)}) \\
& \tilde{\mathbf{A}} = \mathbf{A} + \mathbf{I}, \; \tilde{\mathbf{D}}_{ii} = \sum\nolimits_{j} \tilde{\mathbf{A}}_{ij},  \; \mathbf{H}^{(0)} = \mathbf{X},
\end{aligned}
\end{equation*}
where $\mathbf{A} \in \{0, 1\}^{|V| \times |V|}$ is the adjacent matrix corresponding to the indoor graph $G(V,E)$, $\mathbf{W}^{(\ell)}$ is the learnable weight matrix at the $\ell$-th convolution layer, and $\text{ReLu}(\cdot) = \max(0, \cdot)$ is the activation function.
{GCNs excel at extracting spatial correlations for many tasks, as reported by multiple studies~\cite{kipf2016semi,yu2017spatio,li2017diffusion}.}

The \textit{Temporal-Spatial Unit} is the process to capture the temporal dependencies over the latent vectors extracted by a spatial unit, formally $\mathsf{T}(\mathsf{S}(\mathbf{X}))$.
The representation of its input $\mathbf{X}$ is the same as the spatial unit introduced above.
Likewise, the \textit{Spatial-Temporal Unit} is the process of capturing the spatial dependencies over the latent vectors extracted by a temporal unit, formally $\mathsf{S}(\mathsf{T}(\mathbf{X}))$.
The representation of its input $\mathbf{X}$ is the same as the temporal unit.

As a result, the latent vectors $\mathbf{H}_\textsf{T}$, $\mathbf{H}_\textsf{S}$, $\mathbf{H}_\textsf{TS}$, and $\mathbf{H}_\textsf{ST}$ output by $\textsf{S}(\mathbf{X})$, $\textsf{T}(\mathbf{X})$, $\textsf{T}(\textsf{S}(\mathbf{X}))$, and $\textsf{S}(\textsf{T}(\mathbf{X}))$, respectively, are all in the size of $|V| \times K$.
We feed these latent vectors to a \textit{Self-Attention Unit} to assign proper weights to each latent vector in jointly predicting the result. 
Self-attention is a special type of attention mechanism such that it allows inputs to interact with each other and estimate attention scores (weights) from internal interactions~\cite{vaswani2017attention}.
The formula of our self-attention unit transforms the concatenation $\mathbf{Z}$ of all latent vectors to the attention result $\mathbf{Z}^*$:
\begin{equation*}
\begin{aligned}
& \mathbf{Z}^{*} = \mathbf{V} \cdot \text{softmax}(\mathbf{K}^\mathsf{T} \mathbf{Q}/\sqrt{K''}), \\
& \mathbf{Q} = \mathbf{Z} \mathbf{W}_{Q}, \; \mathbf{K} = \mathbf{Z} \mathbf{W}_{K}, \; \mathbf{V} = \mathbf{Z} \mathbf{W}_{V}, \\
& \mathbf{Z} = [\mathbf{H}_\textsf{T}; \mathbf{H}_\textsf{S}; \mathbf{H}_\textsf{TS}; \mathbf{H}_\textsf{ST}], \\
\end{aligned}
\end{equation*}
where $\mathbf{Q}$, $\mathbf{K}$, $\mathbf{V}$ refer to the queries, keys, and values, respectively, in the self-attention mechanism~\cite{vaswani2017attention}, $\mathbf{W}_{Q}, \mathbf{W}_{K}, \mathbf{W}_{V} \in \mathbb{R}^{K \times K''}$ are their weight matrices accordingly, $K''$ is a dimension of the weight matrices, and $[\cdot; \ldots; \cdot]$ concatenates tensors and transforms the concatenation as a one-dimensional vector.

\smallskip
\noindent\textbf{Output.}
{Given the latent vector $\mathbf{Z}^{*}$ from the self-attention unit, the means $\boldsymbol\mu_{v,t} \in \mathbb{R}^{|V|}$ and standard deviations $\boldsymbol\sigma_{v,t} \in \mathbb{R}^{|V|}$ of all partitions' populations are predicted through an FC layer:}
\begin{equation*}
\boldsymbol\mu_{v,t} = \mathbf{W}_{\mu} \mathbf{Z}^{*} + \mathbf{b}_{\mu}; \;\;\; \boldsymbol\sigma_{v,t} = \mathbf{W}_{\sigma} \mathbf{Z}^{*} + \mathbf{b}_{\sigma}.
\end{equation*}
All matrices $\mathbf{W}_{*}$ and vectors $\mathbf{b}_{*}$ are learnable. The loss and training follow the same idea as SE (see Section~\ref{ssec:local_estimator}).
\section{Estimator-based {CMPP} Processing}
\label{sec:query_processing}

\subsection{Overall Framework}
\label{ssec:framework}

The query processing framework is formalized in Algorithm~\ref{alg:CMPP}.
It receives all parameters of a query instance $\texttt{CMPP}(r, \theta, \eta, t_\text{end},$  $\Delta t)$ and processes the query continuously.

\begin{algorithm}[!htbp]
\footnotesize
\caption{\footnotesize \textsc{CMPP}~(distance $r$, population threshold $\theta$, confidence threshold $\eta$, end time $t_\text{end}$, query time interval $\Delta t$)} \label{alg:CMPP}
\begin{algorithmic}[1]
\State initialize hashtable $\mathcal{H}: V \mapsto (prob, time)$
\State $t_{lq} \gets \varnothing$; $l_l \gets \varnothing$ \Comment{last result updating time; last location}
\While{{current time $t_c$} $\leq t_\text{end}$}
    \State result updating  time $t_q \gets t_c + \Delta t$; obtain query  location $l_q$
    \If{($t_{lq} = \varnothing$ or ${t_q}- t_{lq} \geq \Delta t$) and {$l_q \neq l_l$}}
        \State \textit{result} $\gets$ \Call{Search}{{$l_q$, $r$, $\theta$, $\eta$, $\mathcal{H}$, $t_q$}}
        \State push \textit{result} of $t_q$ to the query user
        \State {$t_{lq} \gets t_q$} \Comment{update last result updating time}
    \EndIf
    \State $l_l \gets$ {$l_q$} \Comment{track location change}
    \State sleep for $1 / z \cdot \Delta t$
\EndWhile
\end{algorithmic}
\end{algorithm}

Initially, a hashtable $\mathcal{H}$ is created to keep a pair of last updated probability $f(P_{v,t} > \theta)$ (see~Definition~\ref{equation:pmf_populated}) and the corresponding update time for each partition (line~1). The information in $\mathcal{H}$ enables lazy computations, i.e., predictions are made only when the current results are no longer valid with respect to a time threshold \texttt{validity}.
The last result updating time $t_{lq}$ and the last query location $l_l$ are also initialized (line~2).
The monitoring is maintained until the {current time} $t_c$ exceeds the end time $t_\text{end}$ (line~3).
{To get near-future monitoring results with the prediction model, a future result updating time $t_q$ is looked ahead as $t_c+\Delta t$ and meanwhile the query location $l_q$ at $t_q$ is obtained.}
%
While monitoring (lines~5--6), the function \textsc{Search}($\cdot$) finds populated partitions when (1) it is the initial time ($t_{lq} = \varnothing$) or {$t_{q}$'s} difference to the last result updating time exceeds the query interval $\Delta t$, and (2) the location just changes ($l_q \neq l_l$). 
When a \textsc{Search} is finished, the returned $\textit{result}$ for time $t_q$ will be pushed to the query user (line~7), and the last result updating time $t_{lq}$ will be renewed.
For each encountered time, the last location $l_l$ is tracked (line~9).
The monitoring is suspended for $1/z \cdot \Delta t$ for reduced workload, where $z$ is a positive integer (line~10).

The \textsc{Search} function is outlined in Algorithm~\ref{alg:search}, expanding outwardly from the query location $l_q$ to all reachable partitions within the range $r$.
In the spirit of Dijkstra's algorithm\footnote{As an alternative, the range-based filtering using $r$ does not speed up the search as it necessitates a further validation of each returned partition according to the indoor distance~\cite{lu2012foundation}.}, this algorithm maintains the distance to $l_q$ and the last-hop partition for each visited door (lines~3--6) and prioritizes the next-door expansion based on a min-heap (lines~7--22).
When a reachable partition $v$ is found (lines~2 and~15), a function \textsc{IsPopulated}($\cdot$) is called. 
If $v$ is tested to be a populated partition, it will be added to the set \textit{result}.

The \textsc{IsPopulated} function (lines~24--28) works as follows.
If the last updated time of $v$ is null or the difference between $t_q$ and $v$'s last updated time exceeds the threshold \texttt{validity}, function \textsc{Population}($\cdot$) is evoked to obtain a new population for $v$ (line~26).
The new population is used to estimate the probability $f(P_{v,t} > \theta)$ (stored as $\mathcal{H}[v].prob$).
If $\mathcal{H}[v].prob \geq \eta$, true is returned to indicate $v$ is populated; otherwise, false is returned (lines~27--28).

\begin{algorithm}[!htbp]
\footnotesize
\caption{\footnotesize \textsc{Search}~(location $l$, distance $r$, population threshold $\theta$, confidence threshold $\eta$, hashtable $\mathcal{H}$, result updating time {$t_q$})} \label{alg:search}
\begin{algorithmic}[1]
\State $\textit{result} \gets \varnothing$; $v \gets \textit{getHostPartition}(l)$
\If{\Call{IsPopulated}{$v$, $\theta$, $\eta$, $t_q$}}
    add $v$ to \textit{result}
\EndIf
\State initialize distance array $\mathit{dist}[]$ for all doors
\State initialize last-hop partition array $\mathit{prev}[]$ for all doors
\For {each door $d \in \mathit{P2D}_\sqsubset(v)$}
	\State $\mathit{dist}[d] \gets ||l, d||_{v}$; $\mathit{prev}[d] \gets v$
\EndFor
\State initialize a min-heap $H$
\For {each door $d_i \in D$}
	\If {$d_i \notin \mathit{P2D}_\sqsubset(v)$}
		$\mathit{dist}[d_i] \gets \infty$; $\mathit{prev}[d_i] \gets$ \textit{null}
	\EndIf

	\State \text{enheap}($H$, $\left \langle d_i, \mathit{dist}[d_i] \right \rangle$); 
	\EndFor

\While {$H$ is not empty}
	\State $\left \langle d_i, \mathit{dist}[d_i] \right \rangle \gets$ \text{deheap}($H$)
	\If {$\mathit{dist}[d_i] > r$}
		\textbf{break}
	\EndIf
	\For {each partition $v_i \in \mathit{D2P}_\sqsupset(d_i)$ $\wedge$ $v_i \neq \mathit{prev}[d_i]$}
	
        \If{\Call{IsPopulated}{$v_i$, $\theta$, $\eta$, {$t_q$}}}
            \State add $v_i$ to \textit{result}
        \EndIf
        
	    \State mark $d_i$ and $v_i$ as visited
    	
    	\For {each unvisited door $d_j \in \mathit{P2D}_\sqsubset(v_i)$}
    		 \State $\mathit{dist}_j \gets \mathit{dist}[d_i] + f_\text{d2d}(v_i,d_i,d_j)$
    		\If {$\mathit{dist}_j < \mathit{dist}[d_j]$}
    			\State $\mathit{dist}[d_j] \gets \mathit{dist}_j$
    			\State \text{enheap}($H$, $\left \langle d_j, \mathit{dist}[d_j] \right \rangle$); $\mathit{prev}[d_j] \gets v_i$
    		\EndIf
    	\EndFor
    	\EndFor
\EndWhile
\State \textbf{return} \textit{result}

\Function{IsPopulated~}{partition $v$, population threshold $\theta$, confidence threshold $\eta$, {result updating time $t_q$}}
\If{$\mathcal{H}[v].\mathit{time}$ is null or $t_q - \mathcal{H}[v].\mathit{time} > $ \texttt{validity}} 
    \State \Call{Population}{$\mathcal{H}, v, \theta, {t_q}$} \Comment{fetch the predicted population}
\EndIf
\If{$\mathcal{H}[v].prob \geq \eta$}
    \textbf{return} true
\EndIf
\State \textbf{else} \textbf{return} false
\EndFunction
\end{algorithmic}
\end{algorithm}

The function \textsc{Population}($\cdot$) is outlined in Algorithm~\ref{alg:test}.
We differentiate two cases, depending on whether a single-way estimator (lines~1--3) or a multi-way estimator (lines~4--8) is prepared.
The two versions are to be detailed in Sections~\ref{ssec:se_predict} and~\ref{ssec:me_predict}, respectively.
{After fetching the population information already predicted for partition $v$ and time $t_q$,}
we call the function \textsc{PMF}($\cdot$) (lines~9--14) to compute the probability $f(P_{v,t} > \theta)$ by looking up the static standard Normal distribution table $\texttt{normTable}$. 
The boundaries for the standard Normal distribution are set to $\pm 4$, {as probabilities corresponding to values out of $[-4, 4]$ are close to zero}.
The probability and the time {$t_q$} will be updated to $\mathcal{H}[v]$ (lines~3 and~8 in Algorithm~\ref{alg:test}).

\begin{algorithm}[!htbp]
\footnotesize
\caption{\footnotesize \textsc{Population}~(hashtable $\mathcal{H}$, partition $v$, population threshold $\theta$, {result updating time $t_q$)}}  \label{alg:test}
\begin{algorithmic}[1]
\If{estimator $\texttt{e}$ is an SE} \Comment{population of one partition}
    \State fetch $(\mu, \sigma)$ {predicted by} \Call{SEPredict}{$\texttt{e}$, $v$, {$t_q$}}
    \State $\mathcal{H}[v] \gets$ (\Call{PMF}{$\mu$, $\sigma$, $\theta$}, {$t_q$})
\EndIf
\If{estimator $\texttt{e}$ is an ME} \Comment{populations of all partitions}
    \State fetch $(\boldsymbol\mu, \boldsymbol\sigma)$ {predicted by} \Call{MEPredict}{$\texttt{e}$, $V$, {$t_q$}}
    \For{each partition $v \in V$}
        \State obtain $(\mu, \sigma)$ for $v$ from $(\boldsymbol\mu, \boldsymbol\sigma)$
        \State $\mathcal{H}[v] \gets$ (\Call{PMF}{$\mu$, $\sigma$, $\theta$}, {$t_q$})
    \EndFor
\EndIf

\Function{PMF~}{mean $\mu$, standard deviation $\sigma$,  threshold $\theta$}
\State $\textit{norm} \gets (\theta - \mu) / \sigma$
\If{$\textit{norm} < -4$}
    \Return 1
\ElsIf{$\textit{norm} < 0$}
    \Return \texttt{normTable}[$-\textit{norm}$] 
\ElsIf{$\textit{norm} \leq 4$}
    \Return 1 - \texttt{normTable}[$\textit{norm}$]
\Else~\Return 0
\EndIf
\EndFunction

\end{algorithmic}
\end{algorithm}

\subsection{\textsc{SEPredict}}
\label{ssec:se_predict}

Algorithm~\ref{alg:se_predict} formalizes \textsc{SEPredict} for an SE.
First, a feature sequence $(\mathbf{x}_1, \ldots,$ $\mathbf{x}_N)$ is generated for $v$ (lines~1--7), where $\mathbf{x}_{i}$ corresponds to timestamp $t_q - (N-i+1)\delta$.
Note that the generation of such a feature sequence online (line~6) is time-consuming due to 
%
%
the probability computations based on Monte Carlo sampling
in Algorithm~\ref{alg:transform}.
Subsequently, the feature sequence is fed into the estimator to make the prediction (line~8).
%
A  caching mechanism for feature sequences is employed to utilize the feature sequence calculated by Algorithm~\ref{alg:transform} in previous timestamps to avoid recomputing from scratch.
Specifically, when the historical population $\mathbf{x}_i$ is generated (line~6), it is stored in a global hashtable \texttt{Cache} (line~7) for potential retrieval in future result updating timestamps (lines~3--4). 

\begin{algorithm}[!htbp]
\footnotesize
\caption{\footnotesize \textsc{SEPredict}~(SE $\texttt{e}$, partition $v$, {result updating time $t_q$)}} \label{alg:se_predict}
\begin{algorithmic}[1]
\For{$i = 1$ to $N$}
    \State $t = {t_q} - (N-i+1) \cdot \delta$
    \If{\texttt{Cache}.hasKey($(v, t)$)} \Comment{\texttt{Cache} is a global resource} %
        \State $\mathbf{x}_i \gets \texttt{Cache}[(v,t)]$
    \Else
        \State $\mathbf{x}_i \gets$ \Call{ExtractPopulation}{$\{ v \}$, $t$} \Comment{Algorithm~\ref{alg:transform}}
        \State $\texttt{Cache}[(v,t)] \gets \mathbf{x}_i$
    \EndIf
\EndFor
\State feed $(\mathbf{x}_1, \ldots, \mathbf{x}_N)$ to $\texttt{e}$ to predict $(\mu, \sigma)$

\end{algorithmic}
\end{algorithm}

\begin{figure}[!htbp]
\centering
\includegraphics[width=\columnwidth]{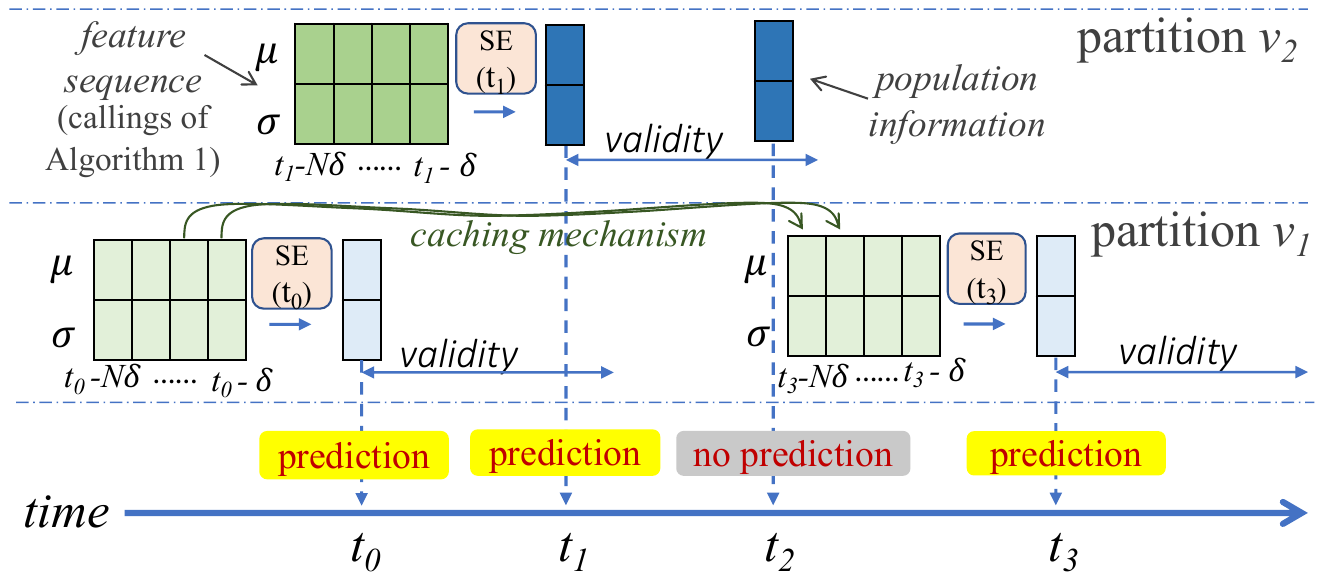}
\caption{{CMPP} processing with SE.}
\label{fig:se_example}
\end{figure}

Fig.~\ref{fig:se_example} illustrates how a {CMPP} is processed over time using an SE. 
Suppose that a {result updating} will happen at a coming time $t_0$, and a reachable partition $v_1$'s population should be predicted for testing if it is populated or not.
To this end, the historical population information $(\mu, \sigma)$ for $v_1$ is obtained at timestamps $\{ t_0 - N\cdot \delta, t_0 - (N-1) \delta, \ldots, t_0 - \delta \}$ based on Algorithm~\ref{alg:transform}. These sequential pairs of mean and variance are fed into SE for predicting $(\mu, \sigma)$ for $v_1$ at $t_0$.
The predicted $(\mu, \sigma)$ at $t_0$ will be valid until time $t_0 + \texttt{validity}$. 

For another partition $v_2$ being tested at time $t_1 < t_0 + \texttt{validity}$, we still need to generate its feature sequence because no prediction was done for $v_2$ between $t_0$ and $t_1$.
However, if $v_2$ needs to be tested at a later time $t_2 < t_1 +\texttt{validity}$, its population at $t_2$ will be approximated as the one predicted at $t_1$ without any computation.
In contrast, $v_1$'s population needs to be predicted at time $t_3 > t_0 + \texttt{validity}$ since $v_1$'s result from $t_0$ is no longer valid at $t_3$.

\noindent\textbf{Remarks}.
\emph{Parts} of the historical population information already computed for $t_0$ can be reused in the feature sequence generation at $t_3$ if there exist some common computation timestamps.
Formally, the condition is $\exists (0< j < i\leq N),~~t_0 - (N-i+1)\delta = 
t_3 - (N-j+1)\delta \Rightarrow t_3 - t_0 = (i - j) \delta$, which strictly requires the difference between $t_3$ and $t_0$ to be an integer multiple of $\delta$ and less than $(N-1) \delta$.

\subsection{\textsc{MEPredict}}
\label{ssec:me_predict}

As an ME predicts populations for all partitions together in one pass, it is more economical to execute such global predictions regularly than the on-demand predictions using SE.
Therefore, \textsc{MEPredict} (Algorithm~\ref{alg:me_predict}) make predictions only at periodic global prediction timestamps (GPTs) that occur with a period of \texttt{validity}.

Algorithm~\ref{alg:me_predict} begins by aligning the timestamp $t_q$ to the most recent GPT $t'$, ensuring $0 \leq t_q - t' < \texttt{validity}$ (line~1). As a result, $t_q$'s population is approximated by $t'$'s.
Lines~2--9 mimic the approach in Algorithm~\ref{alg:se_predict} but generate feature sequences for all partitions.

Generating all partitions' feature sequences will be more time-consuming compared to \textsc{SEPredict}.
Therefore, we stipulate that \texttt{validity} equals $M \cdot \delta$ where $M (< N)$ is a positive integer.
Consequently, the last ($N-M$) populations of previous GPT's feature sequences in the \texttt{Cache} can always be used in the current GPT.
This property makes \textsc{MEPredict}'s caching mechanism highly effective.

\begin{algorithm}[!htbp]
\footnotesize
\caption{\footnotesize \textsc{MEPredict}(ME $\texttt{e}$, partitions $V$, {result updating time $t_q$)}} \label{alg:me_predict}
\begin{algorithmic}[1]
\State $t' = \lfloor ({t_q} / ({M}\delta)) \rfloor \cdot {M}\delta$ \Comment{aligned to the most recent GPT}
\For{$i = 1$ to $N$}
    \State $t = t' - (N-i+1) \cdot \delta$
    \If{\texttt{Cache}.hasKey($(*, t)$)}
        \State $\mathbf{X}_i[v] \gets \texttt{Cache}[(v,t)]$ for each $v \in V$
        \State $\mathbf{X}_i \gets$ combine $\mathbf{X}_i[v]$ of all $v$s
    \Else
        \State $\mathbf{X}_i \gets$ \Call{ExtractPopulation}{$V$, $t$}
        \State $\texttt{Cache}[(v,t)] \gets \mathbf{X}_i[v]$ for each $v \in V$
    \EndIf
\EndFor
\State feed $(\mathbf{X}_1, \ldots, \mathbf{X}_N)$ to $\texttt{e}$ to predict $(\boldsymbol\mu, \boldsymbol\sigma)$

\end{algorithmic}
\end{algorithm}

\begin{figure}[!htbp]
\centering
\includegraphics[width=\columnwidth]{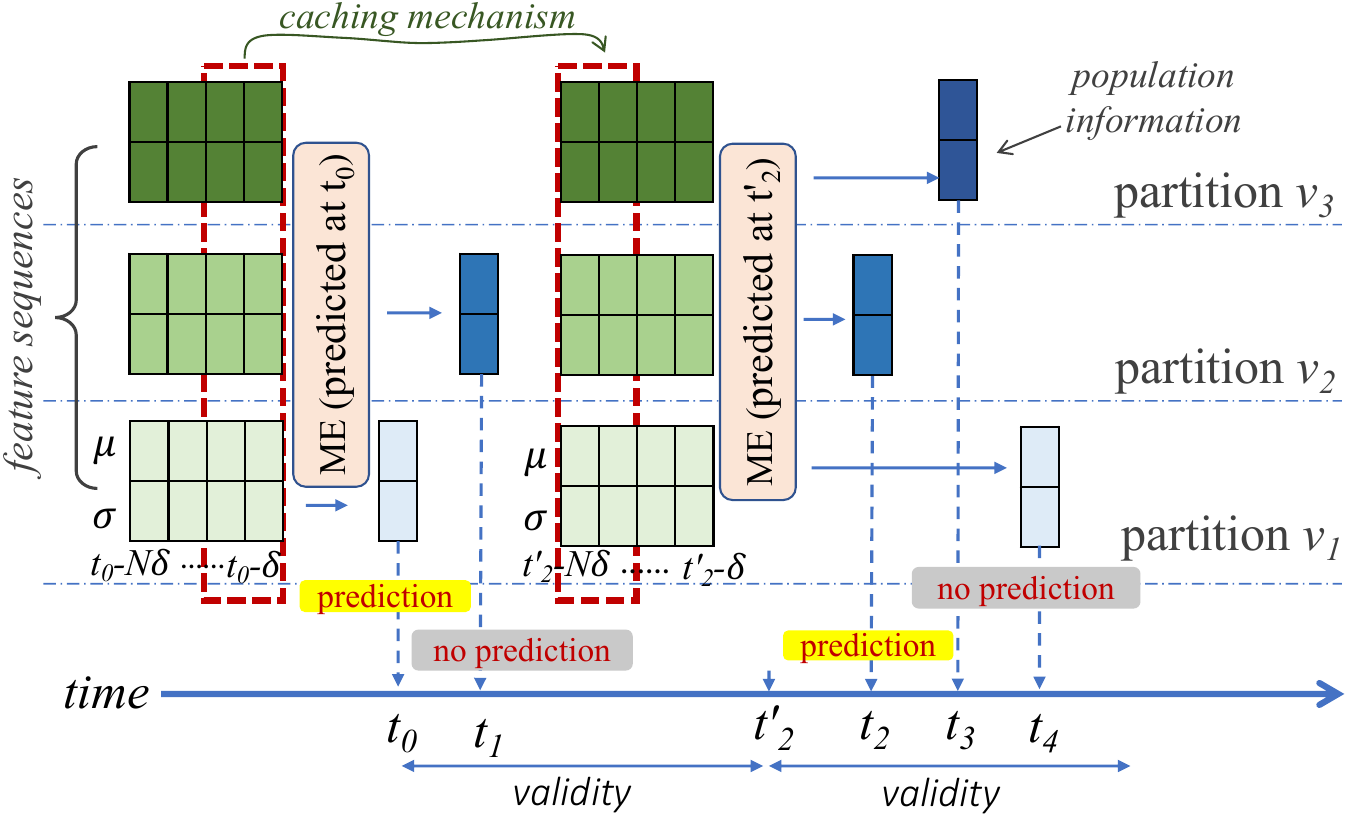}
\caption{{CMPP} processing with ME.}
\label{fig:me_example}
\end{figure}

Fig.~\ref{fig:me_example} illustrates {CMPP} processing using ME.
Suppose $t_0$ is a GPT and a reachable partition $v_1$ needs to be tested at $t_0$.
{All partitions' populations,  including $v_1$'s, will be predicted. 
The predicted results will be valid during the interval $[t_0, t_0 + \texttt{validity})$.}
For another partition $v_2$ that is tested at time $t_1 \in [t_0, t_0 + \texttt{validity})$, its population information is directly retrieved from the results predicted by ME at time $t_0$.
At a later time $t_2$ out of $[t_0, t_0 + \texttt{validity})$, \texttt{MEPredict} is evoked again for testing if partition $v_2$ is populated.
Note that the prediction is done at the most recent GPT $t'_2$ prior to $t_2$ such that the feature sequences cached at $t_0$ can be utilized mostly.
Likewise, all partitions' feature sequences at $t'_2$ are obtained and cached.
In subsequent times $t_3$ and $t_4$, $v_3$'s and $v_1$'s populations are retrieved straightly without predictions since the results at $t'_2$ are still valid.

\vspace*{-5pt}
\section{Experiments}
\label{sec:exp}

\subsection{Experimental Settings}
\label{ssec:exp_settings}

We acquired real indoor positioning data from two buildings (named BLD-1 and BLD-2) {from Jan. 1 to Jan. 31, 2018.}
The data within usual business hours (i.e., 9~a.m.--7~p.m.) was selected for evaluation.
The sampling time intervals in both data sets vary from 5 to 48 seconds with an average of 23 seconds. Table~\ref{tab:datasets} lists key statistics of the used datasets. 

\begin{table}[!htbp]
\footnotesize
\centering
\caption{Data set description.}\label{tab:datasets}
\begin{tabular}{ccc}
\toprule
{Data Set} & {BLD-1} & {BLD-2} \\
\midrule
\# of floors & 7 &6 \\
\# of partitions & 1050  & 900 \\
\# of doors & 1613  &1554 \\
\# of positioning records per day & 91.1k  & 256k\\
\# of trajectories per day & 
1,376  &2,119 \\
\bottomrule
\end{tabular}
\end{table}

For each of the {one-month} data sets, we use the first 22 days (70\%) for training, the next 3 days (10\%) for validation, and the last 6 days (20\%) for population prediction and query processing.
The means and variances for training are all generated by using Algorithm~\ref{alg:transform} on the historical positioning data, {where the number of samplings ($\mathtt{K}$) is set to 200 and the maximum speed ($\mathcal{S}_{max}$) is 1.53 m/s according to an empirical study~\cite{willen2013walking}.}

The estimators are implemented in Python 3.8 + Pytorch 1.8.1, and trained on an NVIDIA GeForce RTX 3080 GPU.
All query processing algorithms are implemented in Java and run on a Mac with a 2.30 GHz Intel i5 CPU and 8 GB memory. 
%
\vspace*{-5pt}
\subsection{Evaluations of Estimators}
\label{ssec:estimators}

\noindent\textbf{Competitor Methods.} 
We compare our SE and ME to the following existing methods that can predict populations.
\begin{enumerate}[leftmargin=*]

    \item \textbf{ARIMA}: Auto-Regressive Integrated Moving Average~\cite{williams2003modeling} is a pure time-series prediction method over variables.
  
    \item \textbf{SVR}: Supported Vector Regression~\cite{awad2015support} is a version of Supported Vector Machine for regression tasks.
    \item \textbf{TGCN}: Temporal Graph Convolutional Network~\cite{zhao2019t} is a stacking of GRU and GCN. Unlike the proposed ME, TGCN uses only the temporal-spatial unit.
    
    \item \textbf{STGCN}: Spatio-Temporal Graph Convolutional Network~\cite{yu2017spatio} is a classical model that combines the GCN (for spatial dependencies) with the gated CNN (for temporal dependencies) to forecast road network traffics.
    \item \textbf{ASTGNN}: Attention based Spatial-Temporal Graph Neural Network~\cite{guo2021learning} is a state-of-the-art model that employs a dynamic GCN with self-attention mechanism.

\end{enumerate}

\smallskip
\noindent\textbf{Model Implementation.}
{We adapt TGCN, STGCN, and ASTGNN to indoor space as follows. First, we change the original distance-based adjacency matrix to a connectivity-based adjacent matrix for graph nodes, each cell of which indicates whether two indoor partitions are connected or not.
Second, we use two separate output layers to predict the mean and variance.
Third, we train the model using the multi-tasking loss function defined in Equation~\ref{equation:wa_loss}.
}
We tune ARIMA and SVR to the best via a parameter grid search.
We set the optimal hyperparameters for TGCN, STGCN, and ASTGNN following their original papers.
For the built-in GRU component in SE and ME, the unit number $N$ is set to 10 and the size of output vector $\mathbf{h}_i$ is set to 16. 
 The number of GCN layers  $\ell$ is set to $2$ and the size of their hidden vector $\mathbf{H}$ is set to 16.
The dimensionality $K''$ is set to 16 for ME's self-attention unit.
Adam optimizer is applied to all networks and the learning rate is set to $0.01$. The iterative training of each neural network stops when the validation loss starts to be flat.
On the same platform, the offline training of ASTGNN, STGCN, TGCN, ME, and SE takes 228, 67.2, 2.9, 8.6, and 2.6 minutes, respectively.
Training of ASTGNN and STGCN is much slower due to their complex internal structures.

\smallskip
\noindent\textbf{Performance Metrics.}
To evaluate the performance of estimators, we employ the Kullback–Leibler (KL) divergence~\cite{kullback1997information}, a typical metric to measure the similarity between two probabilistic distributions. Specifically, we apply the trained estimators on the test data and obtain their estimated distributions, i.e., $\{ \hat{P}_{v,t} = \mathcal{N}(\hat{\mu}_{v,t}, \hat{\sigma}_{v,t}^2) \mid v \in V \}$. Assuming their ground-truth distributions are represented as $\{ P_{v,t} = \mathcal{N}(\mu_{v,t}, \sigma_{v,t}^2) \mid v \in V \}$,
the average KL divergence among partitions is calculated as: 
\begin{equation*}
\text{KL} =  (1/|V|)\sum\nolimits_{v \in V}\  \mathit{D_{KL}(\hat{P}_{v,t} \parallel P_{v,t})},
\end{equation*}
where $\mathit{D_{KL}(\cdot)}$ measures the KL divergence of two distributions. The smaller the above KL value, the more similar the predictions and their ground truth, and thus a better performance of the estimator.

\vspace*{-5pt}
\subsubsection{Effectiveness Study}
\label{sssec:effectiveness}

\begin{table*}[ht]
\footnotesize
\centering
 {\setlength\tabcolsep{4pt} %
\caption{KL vs $\delta$ for all estimators.}\label{tab:kl_vs_delta}
\begin{tabular}{c|ccccccc|ccccccc} 
\toprule
 \multirow{2}{*}{\tabincell{c}{$\delta$\\(min.)}} & \multicolumn{7}{c|}{BLD-1}                         & \multicolumn{7}{c}{BLD-2}                         \\ 
\cmidrule{2-15}
                                       &  ARIMA & SVR & TGCN & STGCN & ASTGNN  & SE     & ME     &  ARIMA & SVR & TGCN & STGCN & ASTGNN  & SE     & ME      \\ 
\midrule
1   & 11.28 & 17.36 & 8.35 & 2.85 & 3.71 & \secBest{1.39} & \best{0.81} &18.01 & 22.53 & 9.56 & 2.56 & 3.75 & \secBest{1.85} & \best{0.95} 

   \\ 
                      5                 & 13.39 & 18.54 & 9.64 & 4.83 & 7.12 & \secBest{2.11} & \best{0.93}    &23.58  & 30.86 & 18.25 & 6.67 & 12.39 & \secBest{2.76} & \best{1.20} 

  \\ 
                      10               & 23.34 & 20.45 & 15.24 & 7.92 & 17.96 & \secBest{3.23} & \best{1.24}     &43.46 & 35.57 & 32.66 & 12.67 & 27.58 & \secBest{3.81} & \best{1.84} 

  \\ 

\bottomrule
\end{tabular}}
\vspace{-7pt}
\end{table*}

Varying the time window $\delta$ of the estimators from {1, 5, and 10 minutes}, we report KL measures over the 6-day test data in Table~\ref{tab:kl_vs_delta}, where the \best{best} and the \secBest{second best} measures for each setting are highlighted. 

Simple prediction methods (i.e., ARIMA and SVR) perform clearly worse than the others, as a pure time-series model fails to capture populations in distinct indoor topological contexts.

Compared to our SE and ME, the outdoor spatiotemporal models (i.e., TGCN, STGCN, and ASTGNN) do not perform well. They rely highly on the distance-dependent adjacency matrix to capture spatial correlations, which however is unavailable in our problem setting. Besides, the characteristics of indoor populations differ from outdoor traffic, which restricts the outdoor models' representation capabilities. In particular, indoor populations are highly uncertain and our models are able to formulate them into probabilistic distributions.
    
Overall, ME outperforms SE in terms of KL in all $\delta$ values. The improvement is attributed to that ME is designed to capture both spatial and temporal dependencies underlying indoor populations, whereas SE only captures temporal dependencies.
Nevertheless, SE still outperforms other competitors, showing its altered GRU architecture is able to learn evolving patterns from indoor populations.
    
Generally, as $\delta$ increases, all methods perform worse. This is natural because long-term predictions are harder. However, ME is robust as it degrades much more slowly than others.

\vspace*{-5pt}
\subsubsection{Persistence Study}
\label{sssec:persistence}

\begin{figure}[!ht]
\centering
\begin{minipage}[t]{0.235\textwidth}
\centering
\includegraphics[width=\textwidth]{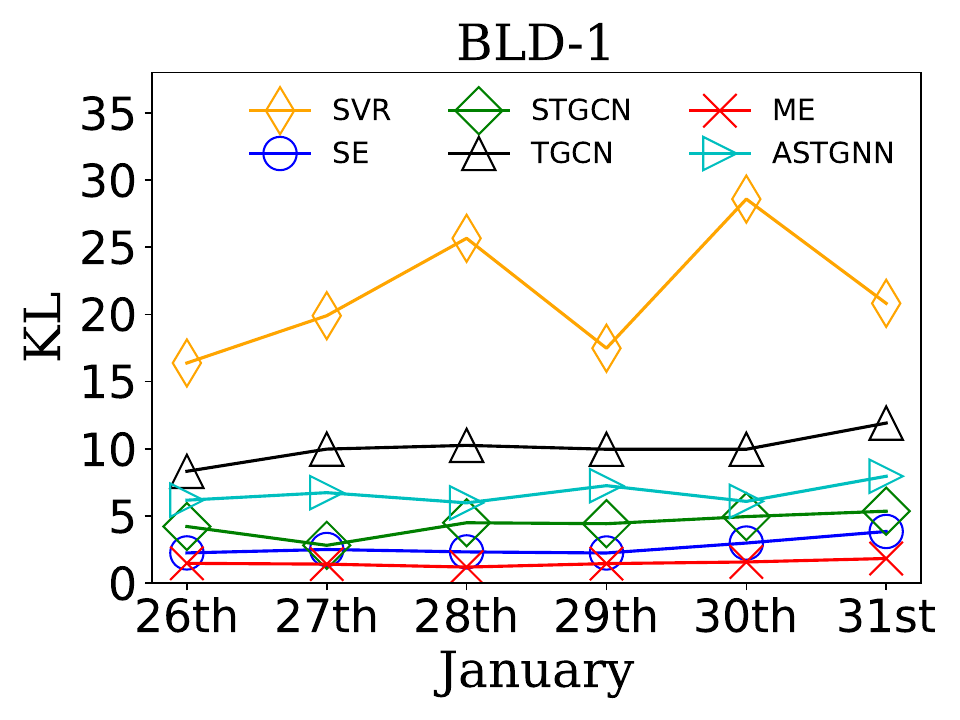}
\ExpCaption{KL vs day (BLD-1).}\label{fig:model_persistance_eu}
\end{minipage}
\begin{minipage}[t]{0.235\textwidth}
\centering
\includegraphics[width=\textwidth]{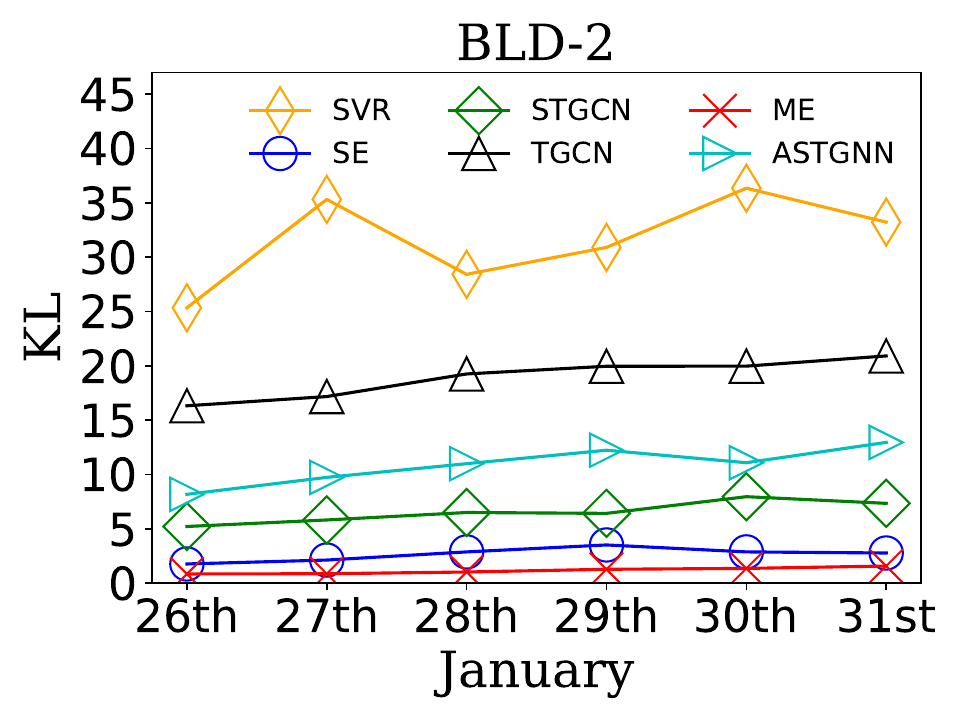}
\ExpCaption{KL vs day (BLD-2).}\label{fig:model_persistance_kl}
\end{minipage}
\end{figure}

We further investigate the persistence of each estimator, i.e., the stability of the prediction accuracy as the trained estimator ages. ARIMA is excluded as it runs in a rolling-forward mode that always needs the most recent data to train the model.
From the six-day test data in BLD-1 and BLD-2, we select each single day to test a trained estimator that is accordingly from one day old until 6 days old. 
We set $\delta=5$ minutes.
The results on BLD-1 and BLD-2 are reported in Figs.~\ref{fig:model_persistance_eu} and~\ref{fig:model_persistance_kl}, respectively.

All models but SVR perform steadily as they become older.
SVR fluctuates widely as it is a simple model and fails to model spatiotemporal dependencies well.
In contrast, all spatiotemporal models based on neural networks are robust over time.
Still, ME and SE outperform others in terms of both measures, showing that our trained estimators work persistently in predicting indoor populations.

\vspace*{-5pt}
\subsubsection{Ablation Study}

\begin{figure}[!ht]
\centering
\begin{minipage}[t]{0.235\textwidth}
\centering
\includegraphics[width=\textwidth]{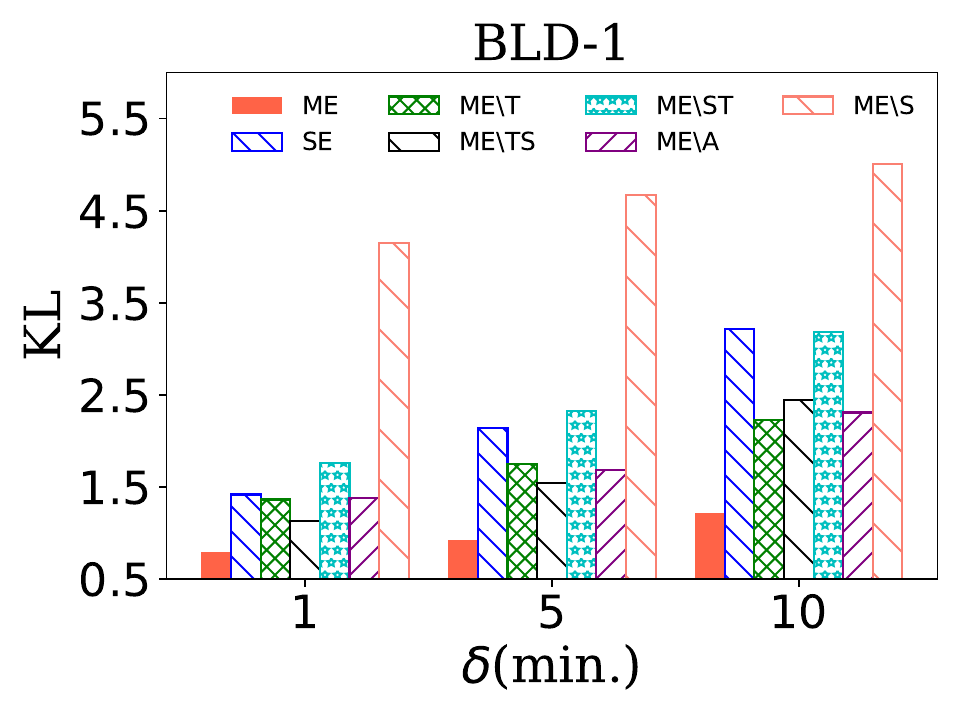}
\ExpCaption{KL vs $\delta$ (BLD-1).}\label{fig:model_ablation_mean}
\end{minipage}
\begin{minipage}[t]{0.235\textwidth}
\centering
\includegraphics[width=\textwidth]{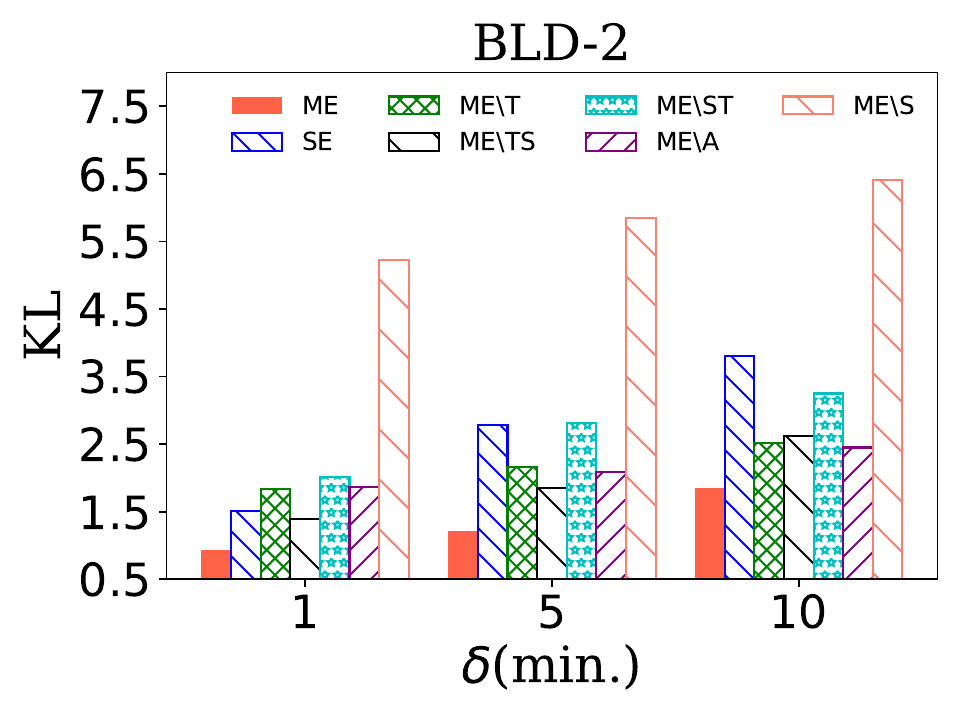}
\ExpCaption{KL vs $\delta$ (BLD-2).}\label{fig:model_ablation_variance}
\end{minipage}
\end{figure}

We investigate the effect of different ME components by excluding them from the original model.
Specifically, ME$\backslash$A excludes the self-attention unit, while
ME$\backslash$T, ME$\backslash$S, ME$\backslash$TS, and ME$\backslash$ST leave out the temporal, spatial, temporal-spatial, and spatial-temporal unit, respectively. The results of BLD-1 and BLD-2 with $\delta=\{1, 5, 10\}$ minutes are reported in Figs.~\ref{fig:model_ablation_mean} and~\ref{fig:model_ablation_variance}.

Compared to the variants, ME generally achieves better results.
This shows that a complete model is able to better extract the regularity from the training data.
ME$\backslash$S performs worst, as it lacks the spatial unit that captures indoor topological correlations. 
Besides, despite their incompleteness, all variants but ME$\backslash$S have comparable performance with SE. This further verifies ME's effectiveness due to its completeness. In our design, apart from the spatial-temporal unit frequently used in outdoor models, the spatial unit, temporal unit, temporal-spatial unit, and attention unit all contribute to the effectiveness of indoor population prediction.

\begin{figure*}[!ht]
\centering
\begin{minipage}[t]{0.23\textwidth}
\centering
\includegraphics[width=\textwidth]{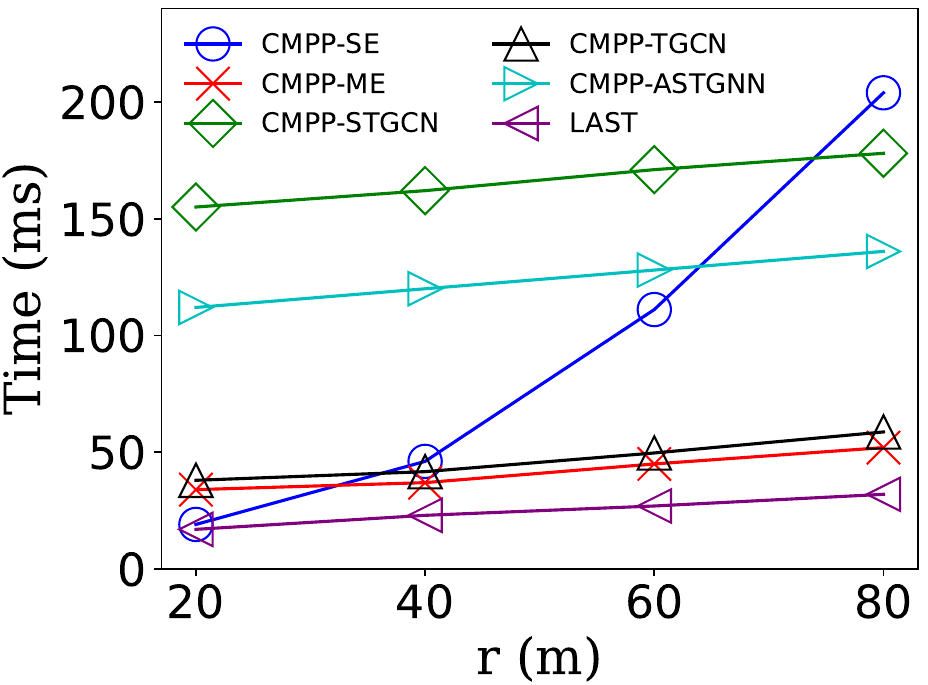}
\ExpCaption{Time vs $r$.}\label{fig:range_time}
\end{minipage}
\begin{minipage}[t]{0.23\textwidth}
\centering
\includegraphics[width=\textwidth]{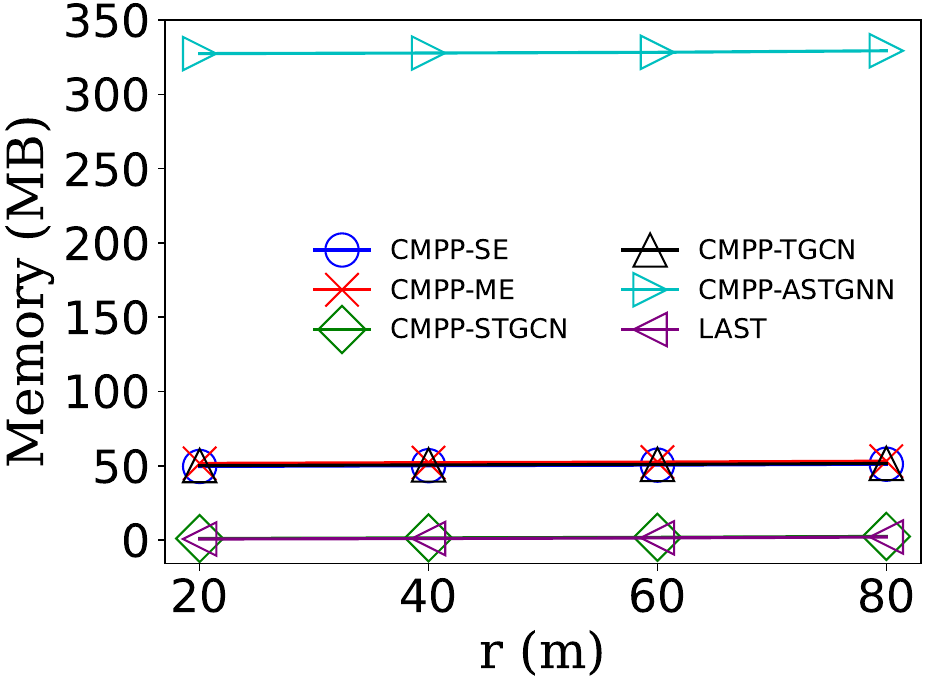}
\ExpCaption{Memory vs $r$.}\label{fig:range_memory}
\end{minipage}
\begin{minipage}[t]{0.23\textwidth}
\centering
\includegraphics[width=\textwidth]{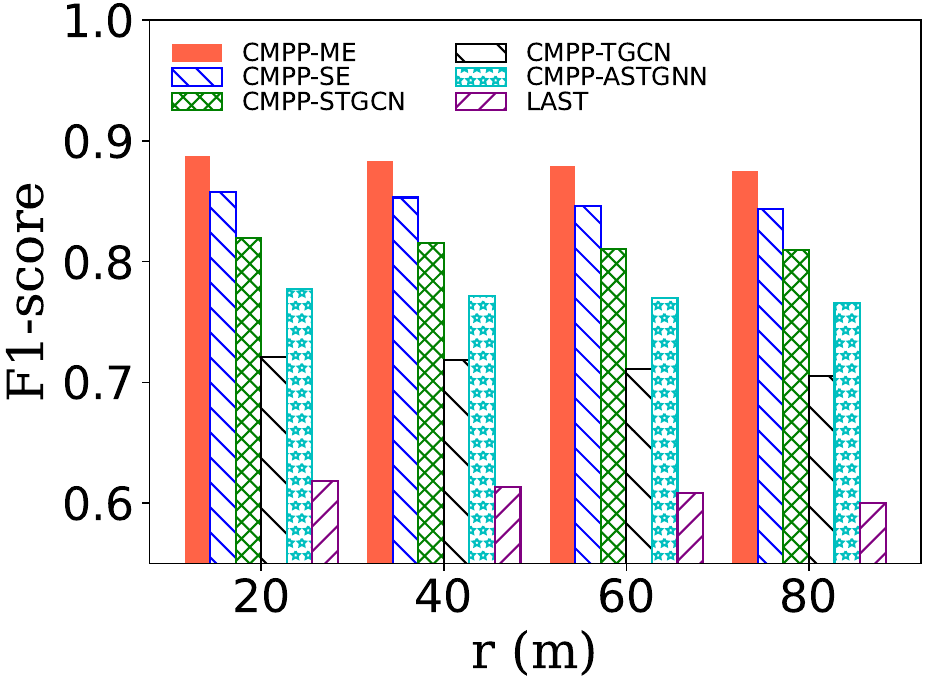}\ExpCaption{F1-score vs $r$.}\label{fig:range_fscore}
\end{minipage}
\begin{minipage}[t]{0.23\textwidth}
\centering
\includegraphics[width=\textwidth]{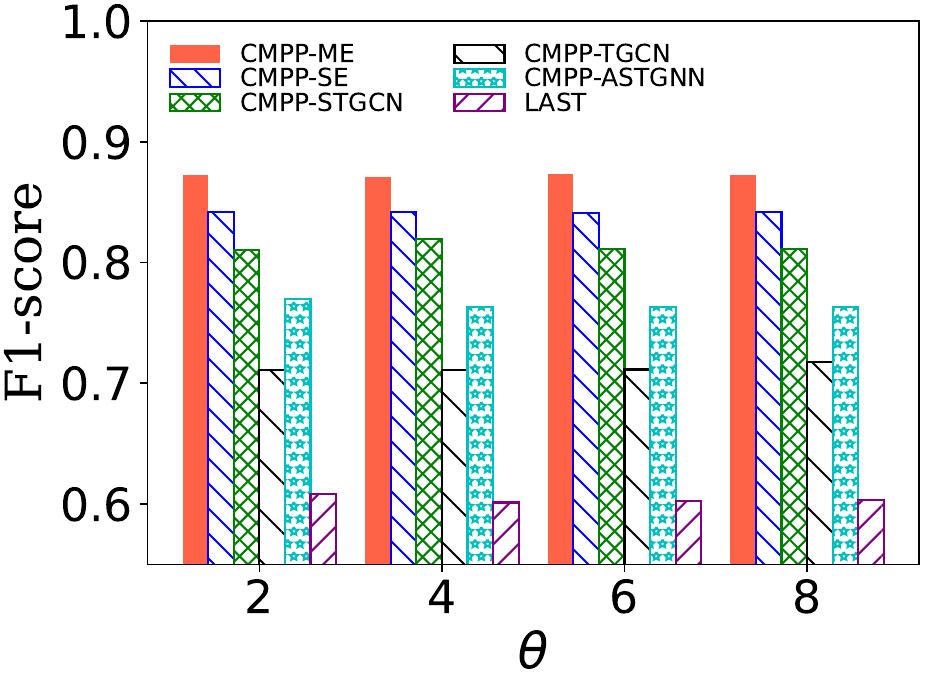}
\ExpCaption{F1-score vs $\theta$.}\label{fig:theta_fscore}
\end{minipage}
\end{figure*}
\begin{figure*}[!ht]
\centering
\begin{minipage}[t]{0.23\textwidth}
\centering
\includegraphics[width=\textwidth]{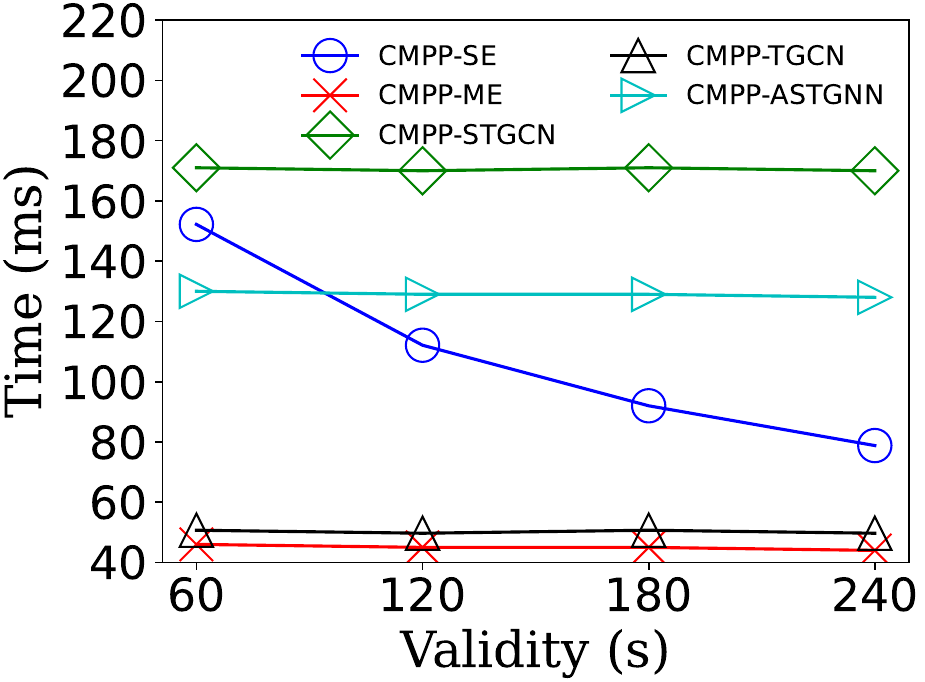}
\ExpCaption{Time vs \texttt{Validity}.}\label{fig:validity_time}
\end{minipage}
\begin{minipage}[t]{0.23\textwidth}
\centering
\includegraphics[width=\textwidth]{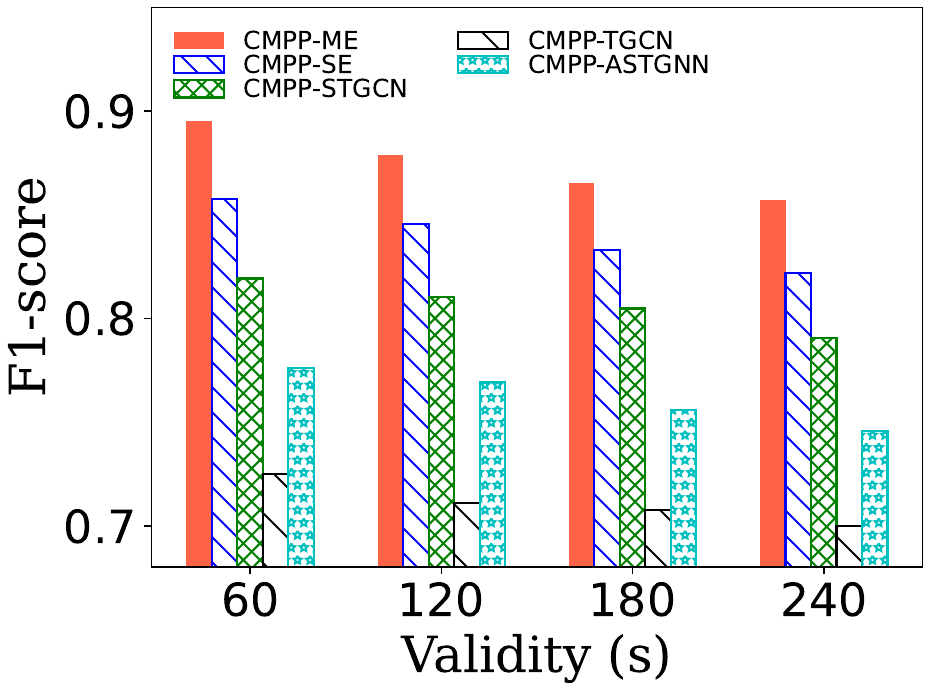}
\ExpCaption{F1-score vs \texttt{Validity}.}\label{fig:validity_fscore}
\end{minipage}
\begin{minipage}[t]{0.23\textwidth}
\centering
\includegraphics[width=\columnwidth]{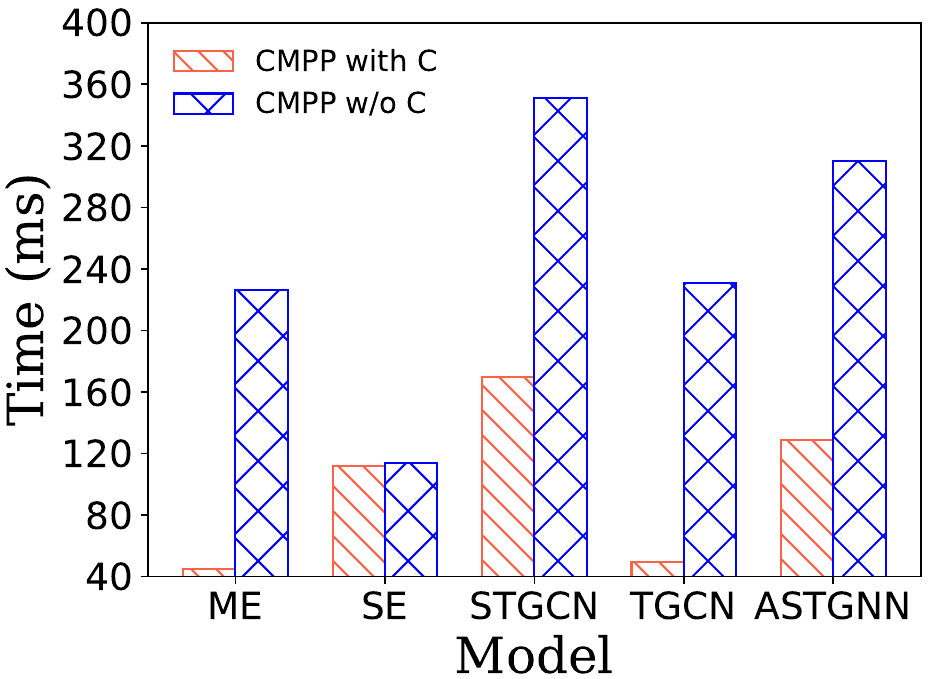}
\ExpCaption{Time vs model.}\label{fig:snapshot_time}
\end{minipage}
\centering
\begin{minipage}[t]{0.23\textwidth}
\centering
\includegraphics[width=\columnwidth]{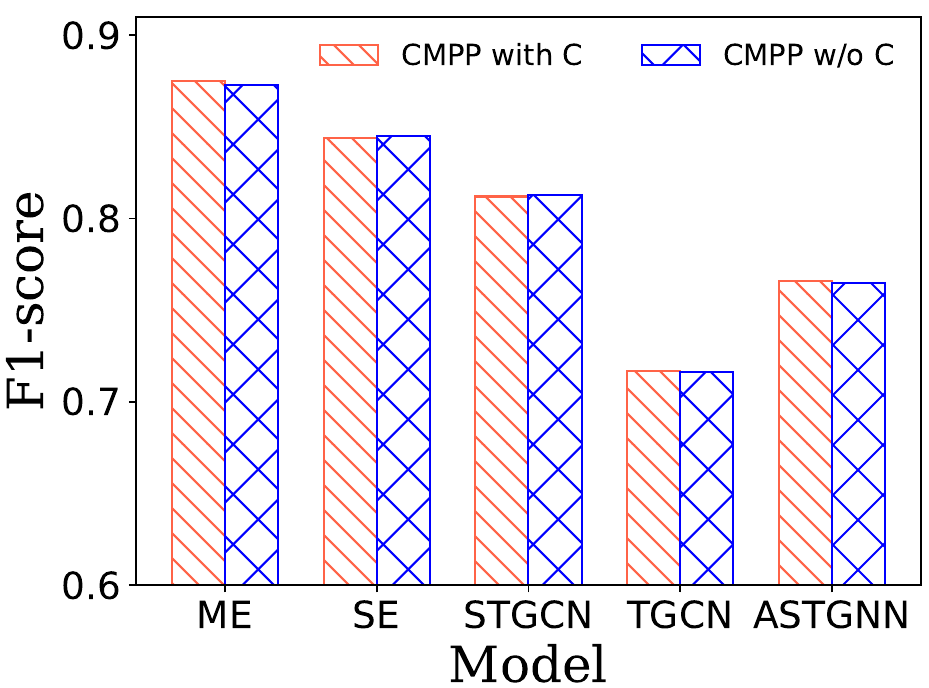}
\ExpCaption{F1-Score vs model.}\label{fig:snapshot_fscore}
\end{minipage}
\vspace*{-10pt}
\end{figure*}

\vspace*{-5pt}
\subsection{Evaluations of {CMPP} Processing}
\label{ssec:query_processing}

We compare our {CMPP} processing algorithms {CMPP-SE} and {CMPP-ME} with three alternatives, namely CMPP-TGCN, CMPP-STGCN, and CMPP-ASTGNN. The alternatives work in the same way as {CMPP-ME} except for a different estimator. {They and {CMPP-ME} are referred to as multi-way estimator-based query processing algorithms, while {CMPP-SE} is single-way estimator-based.} ARIMA and SVR are omitted due to their inferior performance on prediction. 

For CMPP processing, we compare prediction-based methods with a direct counting method, {LAST}, which tallies the last-observed locations of objects in each partition.

\smallskip
\noindent\textbf{Query Instances.}
We simulate 10 indoor instances, each as a user launching a {CMPP} query.
The indoor movement of each object is generated as follows.
First, a dense sequence of indoor locations is randomly decided for an object.
From the first location in the sequence, the object moves 
straight to the next location, stays there for a random period of time from 1 to 500 seconds.
For each object, the query instance lasts for one hour and the query time interval is set to {60 seconds}, the same as $\delta$ of all estimators here.

\smallskip
\noindent\textbf{Performance Metrics.}
For query efficiency, \textbf{response time} is the average time to return the populated partitions, {covering query processing, feature sequences preparation, and inference} for a query timestamp $t$ throughout the continuous query's lifespan. Also, \textbf{memory usage} is measured as the maximum memory consumption for the query lifespan.
For effectiveness, \textbf{F1-score} is used to integrate both precision and recall in finding a list of true populated partitions.
Each metric is the averaged over all 10 query instances.

\smallskip
\noindent\textbf{Query Parameters.}
The parameter settings are listed in Table~\ref{tab:queryparameter} with default values in bold.
We test the effect of each query parameter by varying its value with all others taking their defaults. {The default $\eta$ is set to 0.5.} 
The results for BLD-1 are reported in the rest of this section. Similar results on BLD-2 are reported in the Appendix due to the page limit of the main paper.

\begin{table}[!htbp]
\centering
\caption{Query parameter settings.}\label{tab:queryparameter}
\footnotesize
\begin{tabular}{cc}
\toprule
{Parameter} & {Value}\\
\midrule
$r$ (meter) & {20,40, \textbf{60}, 80} \\
$\theta$ & $\textbf{2}, 4, 6, 8$\\
\texttt{Validity} (second) & 60, $\textbf{120}$, 180, 240\\
\bottomrule
\end{tabular}
\end{table}

\subsubsection{Effect of Query Range $r$}

The query response time and memory usage are reported in Figs.~\ref{fig:range_time} and~\ref{fig:range_memory}, respectively. When $r$ is smaller than $40$ m, {CMPP-SE} incurs less response time than other estimator-based algorithms. 
A smaller $r$ concerns fewer partitions and a single-way estimator can take more advantage of this by only making predictions on fewer partitions. 
In contrast, a multi-way estimator predicts populations on all partitions, although only partial results are needed for small $r$ values.

With $r$ increasing, more reachable partitions need to be considered. As a result, {all the algorithms incur longer response time and consume more memory.
However, {CMPP-SE}'s response time increases much more quickly than others. When more predictions are required to test more partitions,  feature sequences cached by multi-way estimator-based algorithms (i.e., CMPP-ME, CMPP-TGCN, CMPP-STGCN, and CMPP-ASTGNN) are more likely to be reused than the single-way estimator-based method CMPP-SE. When $r$ grows to 80 m, {CMPP-SE} takes the longest response time. The discrepancy among the multi-way estimator-based algorithms is due to their difference in estimators' inference time. ME and TGCN have a relatively simple core network structure and thus infer more quickly. LAST is the fastest since it only gets simple statistics to count populations.}

Regarding memory, {CMPP-ME} uses slightly more than {CMPP-SE}, as {CMPP-ME} on average needs to cache more partitions' information than {CMPP-SE}. Moreover, {CMPP-STGCN} consumes much less memory than {CMPP-ASTGNN}. This is mainly due to the difference in the model size. ASTGNN has an attention-based internal structure that usually needs much more memory than CNN-based STGCN~\cite{shen2018bi}. The model-free LAST uses the least memory.

Referring to the F1-score results reported in  Fig.~\ref{fig:range_fscore}, {CMPP-ME} always outperforms others for different $r$ values.
This is mainly due to that ME is more effective in population prediction than others. In contrast, LAST has the worst performance in finding populated partitions---simply counting recently observed locations is sensitive to the discreteness of indoor positioning data. In addition, the F1-scores of all algorithms decrease slightly with an increasing $r$. We attribute this to that predictions for more partitions possibly make F1-scores reach a slightly lower average. 
\vspace{-6pt}
\subsubsection{Effect of Population Threshold $\theta$}\label{sssec:effect_theta}
As reported in Fig.~\ref{fig:theta_fscore}, varying $\theta$ has almost no effect on the F1-scores of all algorithms since their effectiveness is unrelated to how a populated partition is defined. Likewise, the time and memory measures for different $\theta$ values are all close to their counterparts in the default parameter setting (cf.\ $r=60$ m in Figs.~\ref{fig:range_time} and~\ref{fig:range_memory}). The query parameter $\eta$ also has no effect on the query efficiency and effectiveness; the results are reported in Section~\ref{app:effect_eta} in the Appendix.

\subsubsection{Effect of \texttt{Validity}}
{Referring to Fig.~\ref{fig:validity_time}, the response time decreases quickly for {CMPP-SE} but keeps almost the same for the others as \texttt{Validity} increases. 
A larger \texttt{Validity} renders the population prediction less frequent. 
For {CMPP-SE}, the time cost is reduced accordingly. Nevertheless, for a multi-way estimator-based query processing method like {CMPP-ME} that predicts populations at regular global prediction timestamps (GPTs), a larger \texttt{Validity} means the generated feature sequences for previous GPTs may be reused less for the current GPT. This results in more time cost on preparing feature sequences, which offsets the effect of less frequent predictions. Nevertheless, {CMPP-ME}'s response time remains much lower than {CMPP-SE} even when \texttt{Validity} reaches 240 s.
}
{The validity mechanism has a special effect when \texttt{Validity} equals 60 s, the same as the query interval. In this case, the last valid query result will always expire at the current query time, which invalidates cached results.}

The results of memory usage are omitted here because they are insensitive to \texttt{Validity} in all algorithms.

Fig.~\ref{fig:validity_fscore} shows that all algorithms' F1-scores decrease  with an increasing \texttt{Validity}.
A larger \texttt{Validity} means that the population at a farther future timestamp is allowed to be approximated using the result of last prediction. This reduces the effectiveness of finding populated partitions. Nonetheless, when \texttt{Validity} reaches 240 s, the F1-score for both {CMPP-ME} and {CMPP-SE} still exceeds 0.8. Besides, {CMPP-ME} is always the best in F1-score. We omit LAST as it has no validity mechanism. Its time cost and F1-score are the same as the results of $r = 60$ in Figs.~\ref{fig:range_time} and~\ref{fig:range_fscore}.

\vspace*{-5pt}
\subsubsection{Effect of Caching Mechanism} {
We also investigate the effect of caching mechanism. For each query processing method, we use ``CMPP with C'' and ``CMPP w/o C'' to refer to the version with and without caching, respectively.
As shown in Fig.~\ref{fig:snapshot_time}, {CMPP w/o C} incurs considerably longer response time for all methods. This clearly demonstrates the effect of caching on reducing query time costs. In particular, the time reduction is significant for multi-way estimator-based methods, as multi-way estimators predict populations for all partitions at GPTs and they benefit much more by continuously reusing the cached feature sequences. 
On the contrary, CMPP-SE can only opportunistically reuse the cached feature sequences because it runs in an on-demand way. }

Regarding memory, {CMPP with C} uses about 4 KB more than {CMPP w/o C}, negligible with respect to the total memory cost.

Moreover, the results in Fig.~\ref{fig:snapshot_fscore} show that caching mechanism has no effect on query effectiveness since the goal of caching here is to reuse the historical feature sequences instead of the results of prediction that \texttt{Validity} aims at.

\vspace*{-5pt}
\subsubsection{Summary}
Overall, in various query settings, CMPP-ME performs the best in terms of effectiveness and efficiency. Nevertheless, in the case of smaller $r$s (see $r < 40$ m in Fig.~\ref{fig:range_time}), {CMPP-SE} is more efficient and cost-effective as only few partitions are involved.
LAST, a non-prediction method, runs fast but its effectiveness is rather poor. Also, CMPP-SE benefits more from the validity mechanism, while CMPP-ME takes more advantage of caching.

\vspace*{-7pt}
\section{Related Work}
\label{sec:related}

\begin{table*}[t]
\footnotesize
\centering
 {\setlength\tabcolsep{2pt} %
\caption{Spatiotemporal models for traffic forecasting.}\label{tab:stgcn_related_work}
\begin{tabular}{c|c c c c}
\toprule
Model Name & Spatial Dependency & Temporal Dependency & Architecture & Remarks \\
\midrule
STGCN~\cite{yu2017spatio} & GCN & Gated CNN & \textsf{T}(\textsf{S}) & purely convolutional structures for feature extraction  \\ 
DCRNN~\cite{li2017diffusion}  & Diffusion GCN & Encoder-Decoder & \textsf{T}(\textsf{S}) & diffusion GCN captures bidirectional graph dependencies\\
TGCN~\cite{zhao2019t}  & GCN & GRU & \textsf{T}(\textsf{S}) & GRU has fewer parameters and a simpler structure \\
Graph WaveNet~\cite{wu2019graph} & GCN & Dilated Gated CNN & \textsf{S}(\textsf{T}) & dilated networks handle long-range sequences properly   \\
ASTGCN~\cite{guo2019attention}  & GCN & CNN  & \textsf{T}(\textsf{S}+\textsf{A})+\textsf{A} & attention mechanism for both spatial and temporal modules \\
STSGCN~\cite{song2020spatial}  & GCN & Aggregation & \textsf{T}(\textsf{S}) & synchronous modeling of localized spatial-temporal correlations \\
GMAN~\cite{zheng2020gman}  & Graph Embedding & Encoder-Decoder & \textsf{T}(\textsf{S}+\textsf{A})+\textsf{A} & spatial and temporal attention mechanisms with gated fusion \\
ASTGNN~\cite{guo2021learning}  & Dynamic GCN & Encoder-Decoder & \textsf{T}(\textsf{S}+\textsf{A})+\textsf{A} & dynamic graph with self-attentions \\
\bottomrule
\end{tabular}}
\\ \flushright{\textsf{S}: spatial module; \textsf{T}: temporal module; \textsf{A}: attention mechanism.}
\vspace{-10pt}
\end{table*}

\noindent\textbf{Traffic and Flow Prediction.}
Time-series prediction models such as ARIMA, SVR, and RNN can forecast traffic (or flows)~\cite{ahmed1979analysis}.
However, these models only capture the temporal dependencies among sequential traffic data.
Most recent studies combine spatial and temporal dependencies for predicting traffic on road networks.
For example, spatial dependencies among the traffic data of topologically connected units (i.e., road segments) are modeled by graph convolution networks (GCN)~\cite{yu2017spatio,li2017diffusion,zhao2019t,wu2019graph,guo2019attention,song2020spatial,guo2021learning} or graph embeddings~\cite{zheng2020gman}.
Also, attention~\cite{guo2019attention,zheng2020gman,guo2021learning} can enhance a deep network's capability of identifying key spatial or temporal correlations.
Table~\ref{tab:stgcn_related_work} summarizes and compares such existing works.

These spatiotemporal models fall short in our problem setting.
First, they use a distance-based adjacency matrix to capture the spatial distances between two topological units in spatial dependency modeling, whereas such distances are unavailable for adjacent indoor partitions connected by doors.
Second, they consider traffic as a scalar and thus cannot predict multiple variables jointly, whereas our problem formulation treats indoor populations as probabilistic distributions.
Our ME model uses multi-tasking to predict both mean and variance, and it comprehensively considers four permutations of spatial and temporal units. 
Adapting the existing models\footnote{{We omit DCRNN, Graph WaveNet, ASTGCN, and STSGCN---a previous study~\cite{guo2021learning} shows that they are inferior to ASTGNN. We exclude GMAN since it needs extra embeddings pre-trained from large-scale road networks.}} to our problem fails to achieve good performance (see Section~\ref{sec:exp}).

\smallskip
\noindent\textbf{Querying Populations and Density.}
The estimation of outdoor flows~\cite{tao2004spatio}, populations~\cite{wang2020sclnet}, and object density~\cite{huang2007snapshot, hao2008continuous} has been studied.
However, existing studies do not work in our setting, mainly because 1) indoor positioning data is considerably coarser-grained and sparser than outdoor GPS data, and 2) indoor topology complicates object flows between partitions as spatial units.

Estimating indoor flows and populations has also been studied.
Ahmed et al.~\cite{ahmed2017finding} utilize a graph to extract objects' entry and exit times from raw tracking records in indoor locations.
Li et al.~\cite{li2018finding} count flows of indoor semantic locations during a past time interval by generating possible paths from probabilistic samples.
To compute indoor region density online, Li et al.~\cite{li2018search} construct indoor buffer and core regions to bound the possible movement of an indoor object.
These studies either use RFID-based symbolic tracking data~\cite{ahmed2017finding}, a data type different from what we use in this study or make unique assumptions, e.g., snapshot~\cite{li2018search} or sample-set-based~\cite{li2018finding} positioning records.
As a result, they cannot be applied to modeling the populations from historical location data in our setting. Moreover, they do not generate probabilistic population estimates and thus cannot support answering our CMPP. To the best of our knowledge, we are the first work to model probabilistic partition populations and monitor indoor populations using indoor positioning data  

Some studies~\cite{ma2021wisual,georgievska2019detecting} simply count observed MAC addresses from Wi-Fi signals to estimate indoor crowds. Such methods fall short in counting partition-level populations because they do not consider indoor topology and because a MAC address can be observed in multiple partitions simultaneously. Neither do those studies support real-time monitoring of indoor populations.

To support crowd-aware indoor routing, Liu et al.~\cite{liu2021towards} assume that object flows at doors follow Poisson distributions and estimate the distribution parameters using sequential data in a short period of time. They further explicitly utilize indoor topology to rectify wrong flow estimates.
In contrast, our estimators do not explicitly use indoor topology but learn hidden dependencies from data. 

\smallskip
\noindent\textbf{Continuous Spatial Queries.}
For continuous queries in outdoor spaces, 
safe region~\cite{qi2018continuous} is a key enabling technique. Query reevaluation is needed only if the query location leaves or an object leaves/enters the safe region.
However, safe region techniques fall short in our CMPP processing. As long as a candidate partition's indoor distance to the query location needs to be computed for refinement, the time complexity equals executing a new range query based on Dijkstra's algorithm.
Incremental evaluation based on pruning rules and heuristics~\cite{li2020continuously,zhang2020continuous} is also used in continuous queries.
In contrast, we maintain the validity time period to reduce the number of calls of time-consuming population predictions. 

There are several works on continuous indoor queries.
Yang et al.~\cite{yang2009scalable} study continuous range monitoring queries over uncertain RFID-based symbolic indoor tracking data.
Yuan et al.~\cite{yuan2010supporting} present two network-based approaches to supporting continuous indoor range queries. 
Christensen et al.~\cite{christensen2013continuous} utilize dynamic time- and value-bound semi-constraints to predict object flowing between symbolic locations in processing different spatial queries.
Our work clearly differs from these studies.
First, our populations are extracted based on uncertain geometric positioning data while the studies~\cite{yang2009scalable,christensen2013continuous} retrieve symbolic location data roughly represented by sensor IDs.
Second, our population estimators are fully data-driven and are not based on time or space constraints~\cite{yuan2010supporting,christensen2013continuous}.
Third, our CMPP is a new query type in that it continuously monitors nearby populated partitions for a moving object.

\section{Conclusion and Future Work}
\label{sec:conclusion}

In this work, we model and monitor the populations of indoor partitions with sparse indoor positioning data.  First, we propose a probabilistic method to formulate populations of partitions as Normal distributions. Second, we propose two estimators to predict the on-the-fly population distributions. Based on that, we create a concrete query type that continuously
monitors populated partitions and devise a query processing framework to effectively and efficiently support the query. 
Extensive experiments on real data verify the effectiveness and efficiency of our proposals.

For future work, it is interesting to study other types of continuous queries, e.g., monitoring of the $k$ nearest neighbor populated partitions.



\bibliographystyle{IEEEtran}
\bibliography{uncertain_analysis}

\begin{thebibliography}{10}
\providecommand{\url}[1]{#1}
\csname url@samestyle\endcsname
\providecommand{\newblock}{\relax}
\providecommand{\bibinfo}[2]{#2}
\providecommand{\BIBentrySTDinterwordspacing}{\spaceskip=0pt\relax}
\providecommand{\BIBentryALTinterwordstretchfactor}{4}
\providecommand{\BIBentryALTinterwordspacing}{\spaceskip=\fontdimen2\font plus
\BIBentryALTinterwordstretchfactor\fontdimen3\font minus \fontdimen4\font\relax}
\providecommand{\BIBforeignlanguage}[2]{{%
\expandafter\ifx\csname l@#1\endcsname\relax
\typeout{** WARNING: IEEEtran.bst: No hyphenation pattern has been}%
\typeout{** loaded for the language `#1'. Using the pattern for}%
\typeout{** the default language instead.}%
\else
\language=\csname l@#1\endcsname
\fi
#2}}
\providecommand{\BIBdecl}{\relax}
\BIBdecl

\bibitem{jenkins1992activity}
P.~L. Jenkins, T.~J. Phillips, E.~J. Mulberg, and S.~P. Hui, ``Activity patterns of californians: use of and proximity to indoor pollutant sources,'' \emph{Atmospheric Environment. Part A. General Topics}, vol.~26, no.~12, pp. 2141--2148, 1992.

\bibitem{ott1988human}
W.~R. Ott, \emph{Human activity patterns: A review of the literature for estimating time spent indoors, outdoors, and in transit}.\hskip 1em plus 0.5em minus 0.4em\relax US Environmental Protection Agency, 1988.

\bibitem{cheema2018indoor}
M.~A. Cheema, ``Indoor location-based services: challenges and opportunities,'' \emph{SIGSPATIAL Special}, vol.~10, no.~2, pp. 10--17, 2018.

\bibitem{yaeli2014understanding}
A.~Yaeli, P.~Bak, G.~Feigenblat, S.~Nadler, H.~Roitman, G.~Saadoun, H.~J. Ship, D.~Cohen, O.~Fuchs, S.~Ofek-Koifman \emph{et~al.}, ``Understanding customer behavior using indoor location analysis and visualization,'' \emph{IBM J. Res. Dev.}, vol.~58, no. 5/6, pp. 3--1, 2014.

\bibitem{li2018search}
H.~Li, H.~Lu, L.~Shou, G.~Chen, and K.~Chen, ``In search of indoor dense regions: An approach using indoor positioning data,'' \emph{{IEEE} Trans. Knowl. Data Eng.}, vol.~30, no.~8, pp. 1481--1495, 2018.

\bibitem{world2020covid}
W.~H. Organization \emph{et~al.}, ``Covid-19 management in hotels and other entities of the accommodation sector: interim guidance, 25 august 2020,'' World Health Organization, Tech. Rep., 2020.

\bibitem{liu2007survey}
H.~Liu, H.~Darabi, P.~Banerjee, and J.~Liu, ``Survey of wireless indoor positioning techniques and systems,'' \emph{IEEE Trans. Syst. Man Cybern. Part C}, vol.~37, no.~6, pp. 1067--1080, 2007.

\bibitem{li2023data}
X.~Li, H.~Li, H.~K.-H. Chan, H.~Lu, and C.~S. Jensen, ``Data imputation for sparse radio maps in indoor positioning,'' in \emph{2023 IEEE 39th International Conference on Data Engineering (ICDE)}.\hskip 1em plus 0.5em minus 0.4em\relax IEEE, 2023, pp. 2235--2248.

\bibitem{wang2019fast}
S.~Wang, Z.~Bao, J.~S. Culpepper, T.~Sellis, and X.~Qin, ``Fast large-scale trajectory clustering,'' \emph{PVLDB}, vol.~13, no.~1, pp. 29--42, 2019.

\bibitem{tang2019joint}
X.~Tang, B.~Gong, Y.~Yu, H.~Yao, Y.~Li, H.~Xie, and X.~Wang, ``Joint modeling of dense and incomplete trajectories for citywide traffic volume inference,'' in \emph{WWW}, 2019, pp. 1806--1817.

\bibitem{zheng2012reducing}
K.~Zheng, Y.~Zheng, X.~Xie, and X.~Zhou, ``Reducing uncertainty of low-sampling-rate trajectories,'' in \emph{ICDE}, 2012, pp. 1144--1155.

\bibitem{wu2016probabilistic}
H.~Wu, J.~Mao, W.~Sun, B.~Zheng, H.~Zhang, Z.~Chen, and W.~Wang, ``Probabilistic robust route recovery with spatio-temporal dynamics,'' in \emph{KDD}, 2016, pp. 1915--1924.

\bibitem{jensen2009graph}
C.~S. Jensen, H.~Lu, and B.~Yang, ``Graph model based indoor tracking,'' in \emph{MDM}, 2009, pp. 122--131.

\bibitem{lu2012foundation}
H.~Lu, X.~Cao, and C.~S. Jensen, ``A foundation for efficient indoor distance-aware query processing,'' in \emph{ICDE}, 2012, pp. 438--449.

\bibitem{liu2021towards}
T.~Liu, H.~Li, H.~Lu, M.~A. Cheema, and L.~Shou, ``Towards crowd-aware indoor path planning,'' \emph{PVLDB}, vol.~14, no.~8, pp. 1365--1377, 2021.

\bibitem{inam2018safety}
R.~Inam, E.~Fersman, K.~Raizer, R.~Souza, A.~Nascimento, and A.~Hata, ``Safety for automated warehouse exhibiting collaborative robots,'' in \emph{Safety and Reliability--Safe Societies in a Changing World}.\hskip 1em plus 0.5em minus 0.4em\relax CRC Press, 2018, pp. 2021--2028.

\bibitem{DBLP:conf/edbt/ZhangPMZ04}
J.~Zhang, D.~Papadias, K.~Mouratidis, and M.~Zhu, ``Spatial queries in the presence of obstacles,'' in \emph{{EDBT}}, 2004, pp. 366--384.

\bibitem{wang1993number}
Y.~H. Wang, ``On the number of successes in independent trials,'' \emph{Statistica Sinica}, pp. 295--312, 1993.

\bibitem{fernandez2010closed}
M.~Fern{\'a}ndez and S.~Williams, ``Closed-form expression for the poisson-binomial probability density function,'' \emph{IEEE Trans. Aerosp. Electron. Syst.}, vol.~46, no.~2, pp. 803--817, 2010.

\bibitem{hogg1977probability}
R.~V. Hogg, E.~A. Tanis, and D.~L. Zimmerman, \emph{Probability and statistical inference}.\hskip 1em plus 0.5em minus 0.4em\relax Macmillan New York, 1977, vol. 993.

\bibitem{neammanee2005refinement}
K.~Neammanee, ``A refinement of normal approximation to poisson binomial,'' \emph{Int. J. Math. Sci.}, vol. 2005, no.~5, pp. 717--728, 2005.

\bibitem{willen2013walking}
C.~Willen, K.~Lehmann, and K.~Sunnerhagen, ``Walking speed indoors and outdoors in healthy persons and in persons with late effects of polio,'' \emph{J. Neurosci. Res.}, vol.~3, no.~2, pp. 62--67, 2013.

\bibitem{yawalkar2019route}
P.~Yawalkar and S.~Ranu, ``Route recommendations on road networks for arbitrary user preference functions,'' in \emph{ICDE}, 2019, pp. 602--613.

\bibitem{chung2014empirical}
J.~Chung, C.~Gulcehre, K.~Cho, and Y.~Bengio, ``Empirical evaluation of gated recurrent neural networks on sequence modeling,'' in \emph{NeurIPS Workshop on Deep Learning}, 2014.

\bibitem{zhao2019t}
L.~Zhao, Y.~Song, C.~Zhang, Y.~Liu, P.~Wang, T.~Lin, M.~Deng, and H.~Li, ``{T-GCN}: A temporal graph convolutional network for traffic prediction,'' \emph{IEEE Trans. Intell. Transp. Syst.}, vol.~21, no.~9, pp. 3848--3858, 2019.

\bibitem{hettige2020robust}
B.~Hettige, W.~Wang, Y.-F. Li, and W.~Buntine, ``Robust attribute and structure preserving graph embedding,'' in \emph{Advances in Knowledge Discovery and Data Mining: 24th Pacific-Asia Conference, PAKDD 2020, Singapore, May 11--14, 2020, Proceedings, Part II 24}.\hskip 1em plus 0.5em minus 0.4em\relax Springer, 2020, pp. 593--606.

\bibitem{kipf2016semi}
T.~N. Kipf and M.~Welling, ``Semi-supervised classification with graph convolutional networks,'' in \emph{ICLR}, 2017.

\bibitem{yu2017spatio}
B.~Yu, H.~Yin, and Z.~Zhu, ``Spatio-temporal graph convolutional networks: A deep learning framework for traffic forecasting,'' in \emph{IJCAI}, 2018, pp. 3634--3640.

\bibitem{li2017diffusion}
Y.~Li, R.~Yu, C.~Shahabi, and Y.~Liu, ``Diffusion convolutional recurrent neural network: Data-driven traffic forecasting,'' in \emph{ICLR}, 2018.

\bibitem{vaswani2017attention}
A.~Vaswani, N.~Shazeer, N.~Parmar, J.~Uszkoreit, L.~Jones, A.~N. Gomez, L.~Kaiser, and I.~Polosukhin, ``Attention is all you need,'' pp. 5998--6008, 2017.

\bibitem{williams2003modeling}
B.~M. Williams and L.~A. Hoel, ``Modeling and forecasting vehicular traffic flow as a seasonal {ARIMA} process: Theoretical basis and empirical results,'' \emph{Journal of Transportation Engineering}, vol. 129, no.~6, pp. 664--672, 2003.

\bibitem{awad2015support}
M.~Awad and R.~Khanna, ``Support vector regression,'' in \emph{Efficient learning machines}.\hskip 1em plus 0.5em minus 0.4em\relax Springer, 2015, pp. 67--80.

\bibitem{guo2021learning}
S.~Guo, Y.~Lin, H.~Wan, X.~Li, and G.~Cong, ``Learning dynamics and heterogeneity of spatial-temporal graph data for traffic forecasting,'' \emph{{IEEE} Trans. Knowl. Data Eng.}, vol.~34, no.~11, pp. 5415--5428, 2022.

\bibitem{kullback1997information}
S.~Kullback, \emph{Information theory and statistics}.\hskip 1em plus 0.5em minus 0.4em\relax Courier Corporation, 1997.

\bibitem{shen2018bi}
T.~Shen, T.~Zhou, G.~Long, J.~Jiang, and C.~Zhang, ``Bi-directional block self-attention for fast and memory-efficient sequence modeling,'' in \emph{ICLR}, 2018.

\bibitem{wu2019graph}
Z.~Wu, S.~Pan, G.~Long, J.~Jiang, and C.~Zhang, ``Graph wavenet for deep spatial-temporal graph modeling,'' in \emph{IJCAI}, 2019, pp. 1907--1913.

\bibitem{guo2019attention}
S.~Guo, Y.~Lin, N.~Feng, C.~Song, and H.~Wan, ``Attention based spatial-temporal graph convolutional networks for traffic flow forecasting,'' in \emph{AAAI}, vol.~33, no.~01, 2019, pp. 922--929.

\bibitem{song2020spatial}
C.~Song, Y.~Lin, S.~Guo, and H.~Wan, ``Spatial-temporal synchronous graph convolutional networks: A new framework for spatial-temporal network data forecasting,'' in \emph{AAAI}, vol.~34, no.~01, 2020, pp. 914--921.

\bibitem{zheng2020gman}
C.~Zheng, X.~Fan, C.~Wang, and J.~Qi, ``{GMAN}: A graph multi-attention network for traffic prediction,'' in \emph{AAAI}, vol.~34, no.~01, 2020, pp. 1234--1241.

\bibitem{ahmed1979analysis}
M.~S. Ahmed and A.~R. Cook, \emph{Analysis of freeway traffic time-series data by using {Box-Jenkins} techniques}.\hskip 1em plus 0.5em minus 0.4em\relax Transportation Research Board, 1979, no. 722.

\bibitem{tao2004spatio}
Y.~Tao, G.~Kollios, J.~Considine, F.~Li, and D.~Papadias, ``Spatio-temporal aggregation using sketches,'' in \emph{ICDE}, 2004, pp. 214--225.

\bibitem{wang2020sclnet}
S.~Wang, Y.~Lu, T.~Zhou, H.~Di, L.~Lu, and L.~Zhang, ``{SCLNet}: Spatial context learning network for congested crowd counting,'' \emph{Neurocomputing}, vol. 404, pp. 227--239, 2020.

\bibitem{huang2007snapshot}
X.~Huang and H.~Lu, ``Snapshot density queries on location sensors,'' in \emph{SIGMOD MobiDE}, 2007, pp. 75--78.

\bibitem{hao2008continuous}
X.~Hao, X.~Meng, and J.~Xu, ``Continuous density queries for moving objects,'' in \emph{SIGMOD MobiDE}, 2008, pp. 1--7.

\bibitem{ahmed2017finding}
T.~Ahmed, T.~B. Pedersen, and H.~Lu, ``Finding dense locations in symbolic indoor tracking data: Modeling, indexing, and processing,'' \emph{GeoInformatica}, vol.~21, no.~1, pp. 119--150, 2017.

\bibitem{li2018finding}
H.~Li, H.~Lu, L.~Shou, G.~Chen, and K.~Chen, ``Finding most popular indoor semantic locations using uncertain mobility data,'' \emph{{IEEE} Trans. Knowl. Data Eng.}, vol.~31, no.~11, pp. 2108--2123, 2018.

\bibitem{ma2021wisual}
X.~Ma, W.~Xi, X.~Zhao, Z.~Chen, H.~Zhang, and J.~Zhao, ``Wisual: Indoor crowd density estimation and distribution visualization using wi-fi,'' \emph{IEEE Internet of Things Journal}, vol.~9, no.~12, pp. 10\,077--10\,092, 2021.

\bibitem{georgievska2019detecting}
S.~Georgievska, P.~Rutten, J.~Amoraal, E.~Ranguelova, R.~Bakhshi, B.~L. de~Vries, M.~Lees, and S.~Klous, ``Detecting high indoor crowd density with wi-fi localization: A statistical mechanics approach,'' \emph{Journal of Big Data}, vol.~6, no.~1, pp. 1--23, 2019.

\bibitem{qi2018continuous}
J.~Qi, R.~Zhang, C.~S. Jensen, K.~Ramamohanarao, and J.~He, ``Continuous spatial query processing: A survey of safe region based techniques,'' \emph{ACM Computing Surveys}, vol.~51, no.~3, pp. 1--39, 2018.

\bibitem{li2020continuously}
L.~Li, M.~A. Cheema, M.~E. Ali, H.~Lu, and D.~Taniar, ``Continuously monitoring alternative shortest paths on road networks,'' \emph{PVLDB}, vol.~13, no.~12, pp. 2243--2255, 2020.

\bibitem{zhang2020continuous}
D.~Zhang, Z.~Chang, S.~Wu, Y.~Yuan, K.-L. Tan, and G.~Chen, ``Continuous trajectory similarity search for online outlier detection,'' \emph{{IEEE} Trans. Knowl. Data Eng.}, vol.~34, no.~10, pp. 4690--4704, 2020.

\bibitem{yang2009scalable}
B.~Yang, H.~Lu, and C.~S. Jensen, ``Scalable continuous range monitoring of moving objects in symbolic indoor space,'' in \emph{CIKM}, 2009, pp. 671--680.

\bibitem{yuan2010supporting}
W.~Yuan and M.~Schneider, ``Supporting continuous range queries in indoor space,'' in \emph{MDM}, 2010, pp. 209--214.

\bibitem{christensen2013continuous}
K.~F. Christensen, L.~L. Christiansen, T.~B. Pedersen, and J.~Pihl, ``Continuous query processing for actual and predicted object flow in symbolic space,'' in \emph{MDM}, vol.~1, 2013, pp. 217--226.

\end{thebibliography}

\clearpage


\appendices

\section{Evaluations of {CMPP} Processing}
\label{ssec:query_processing}

The supplementary experiments consist of the experiments of the whole querying processing for BLD-2 (i.e., Figs.~\ref{fig:range_time}--\ref{fig:eta_memo} and Figs. \ref{fig:duration_time}--\ref{fig:snapshot_fscore}), and the experiments of $\eta$'s effect for BLD-1 (i.e., Figs. \ref{fig:b1_eta_time}--\ref{fig:b1_eta_memo}).
\begin{table}[!htbp]
\centering
\caption{Query parameter settings.}\label{tab:queryparameter}
\small
\begin{tabular}{|c|c|}
\hline
\textbf{Parameter} & \textbf{Value}\\
\hline
$r$ (meter) & {20,40, \textbf{60}, 80} \\
\hline
$\theta$ & $\textbf{2}, 4, 6, 8$\\
\hline
\texttt{Validity} (second) & 60, $\textbf{120}$, 180, 240\\
\hline
$\eta$ &  $\textbf{0.5}$, 0.6, 0.7, 0.8\\
\hline
\end{tabular}
\end{table}

\noindent\textbf{Query Instance.}
We generate query instances in the same way as that in the paper.

\noindent\textbf{Performance Metrics.}
We use the same performance metrics as that in the paper.

\noindent\textbf{Query Parameters.}
Besides the query parameters (i.e., $r$, $\theta$, and \texttt{Validity})  in the paper, we supply the experimental results of $\eta$'s effect. 


\subsection{Effect of Query Range $r$.}\label{app:effect_range}
\begin{figure*}[!ht]
\centering
\begin{minipage}[t]{0.23\textwidth}
\centering
\includegraphics[width=\textwidth]{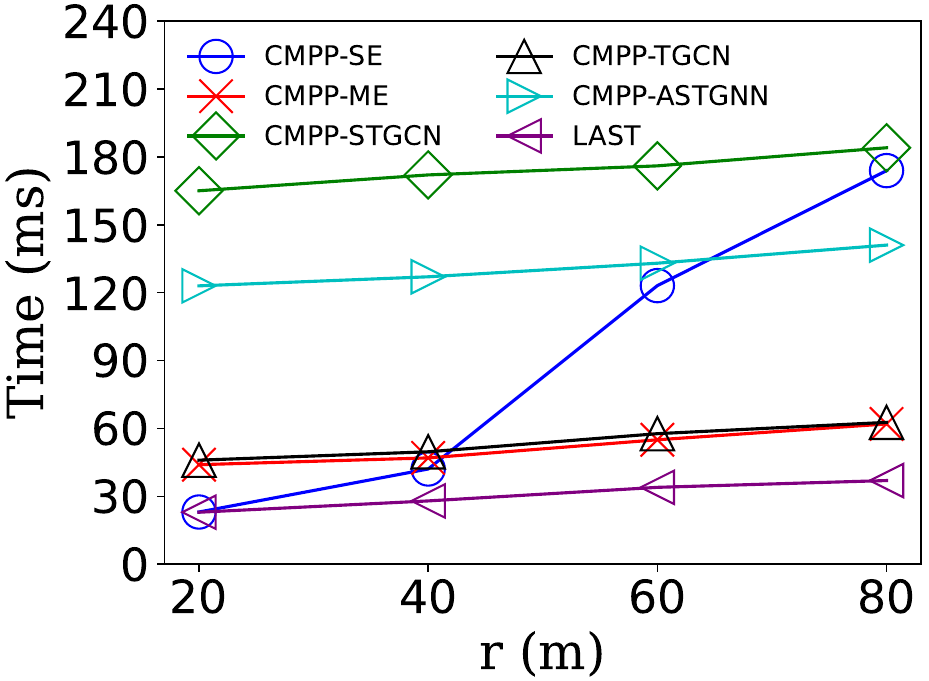}
\ExpCaption{Time vs $r$ (BLD-2).}\label{fig:range_time}
\end{minipage}
\begin{minipage}[t]{0.23\textwidth}
\centering
\includegraphics[width=\textwidth]{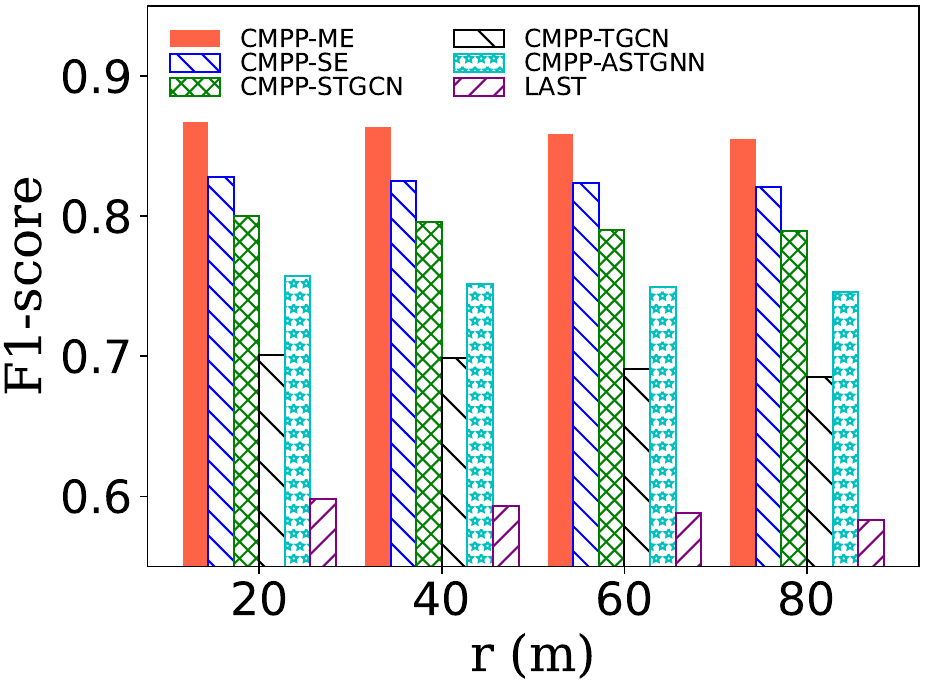}
\ExpCaption{F1-score vs $r$ (BLD-2).}\label{fig:range_fscore}
\end{minipage}
\begin{minipage}[t]{0.23\textwidth}
\centering
\includegraphics[width=\textwidth]{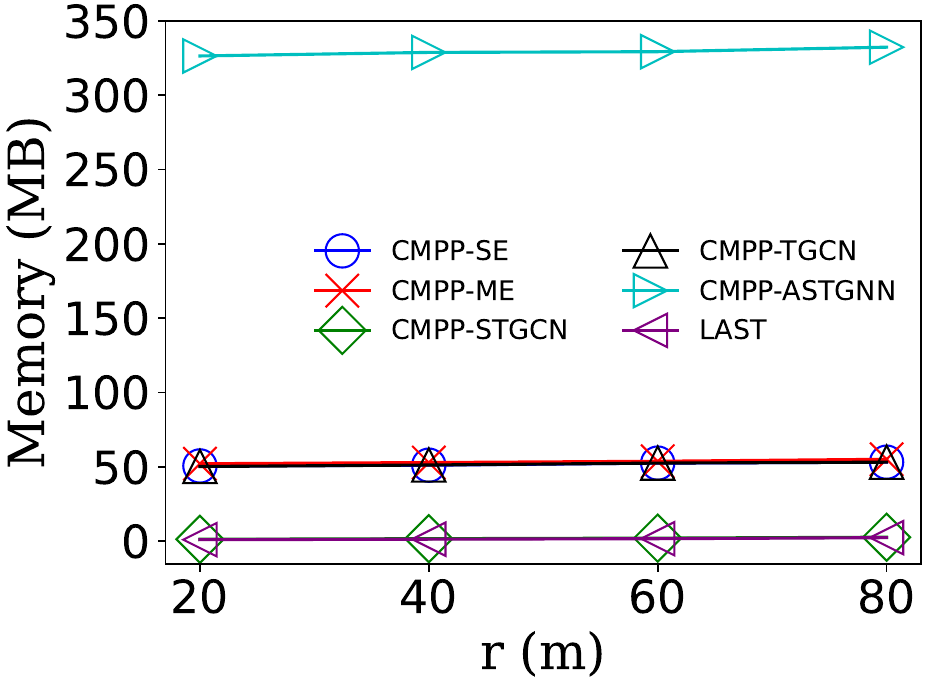}
\ExpCaption{Memory vs $r$ (BLD-2).}\label{fig:range_memo}
\end{minipage}
\end{figure*}

The query response time, F1-score, and memory usage for BLD-2 are reported in Fig.~\ref{fig:range_time}, Fig.~\ref{fig:range_fscore}, and Fig.~\ref{fig:range_memo}, respectively. They follow similar tendencies as that in BLD-1 (cf. Section 6.3.1 in the paper) and detailed interpretation could be found there.

\subsection{Effect of Population Threshold $\theta$.}\label{app:effect_thre} 
\begin{figure*}[!ht]
\centering
\begin{minipage}[t]{0.23\textwidth}
\centering
\includegraphics[width=\textwidth]{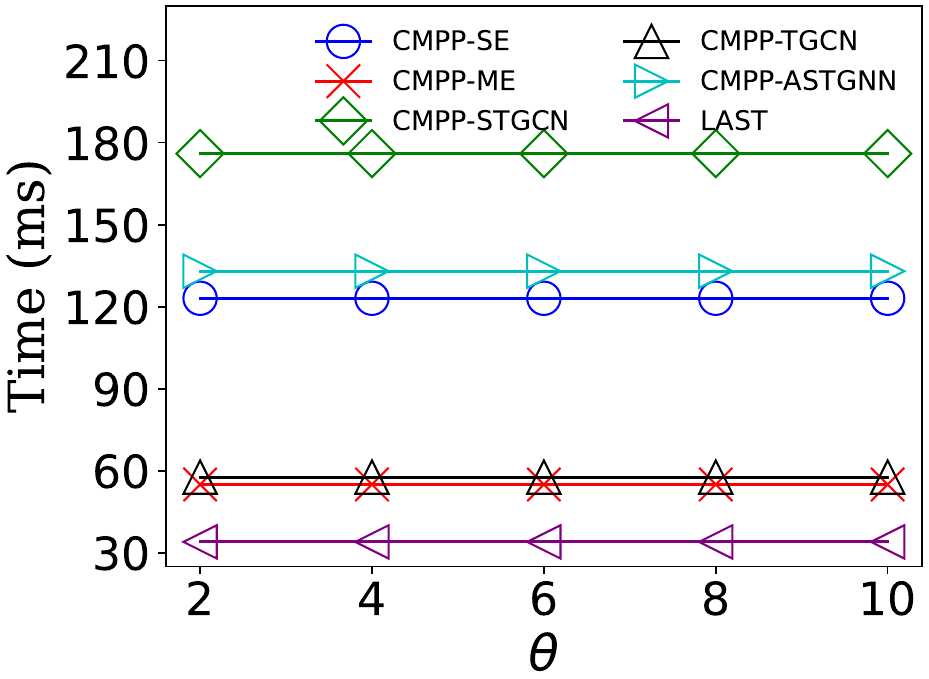}
\ExpCaption{Time vs $\theta$ (BLD-2).}\label{fig:popu_time}
\end{minipage}
\begin{minipage}[t]{0.23\textwidth}
\centering
\includegraphics[width=\textwidth]{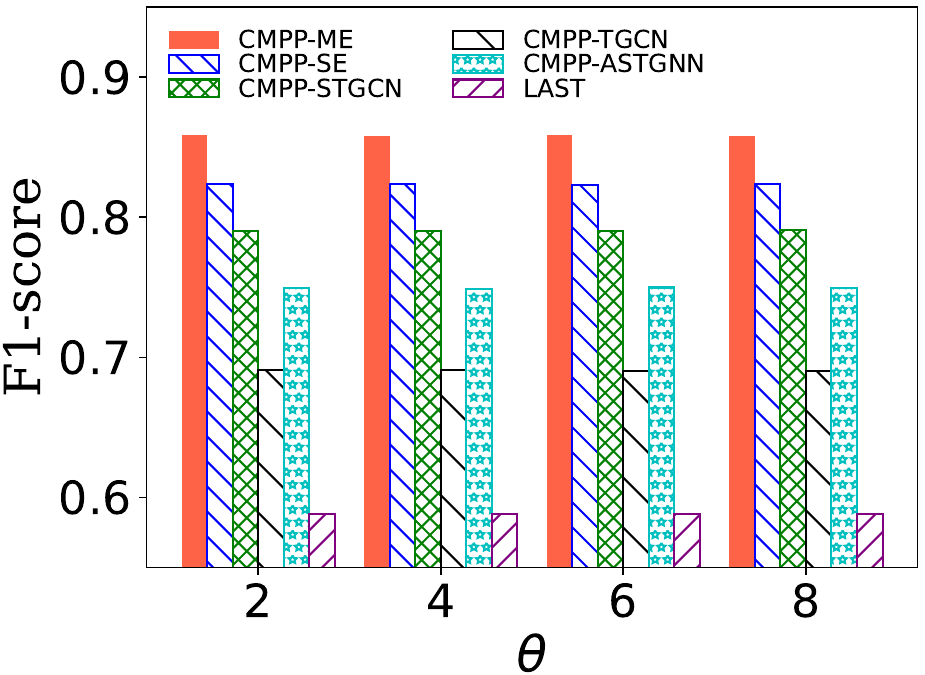}
\ExpCaption{F1-score vs $\theta$ (BLD-2).}\label{fig:popu_fscore}
\end{minipage}
\begin{minipage}[t]{0.23\textwidth}
\centering
\includegraphics[width=\textwidth]{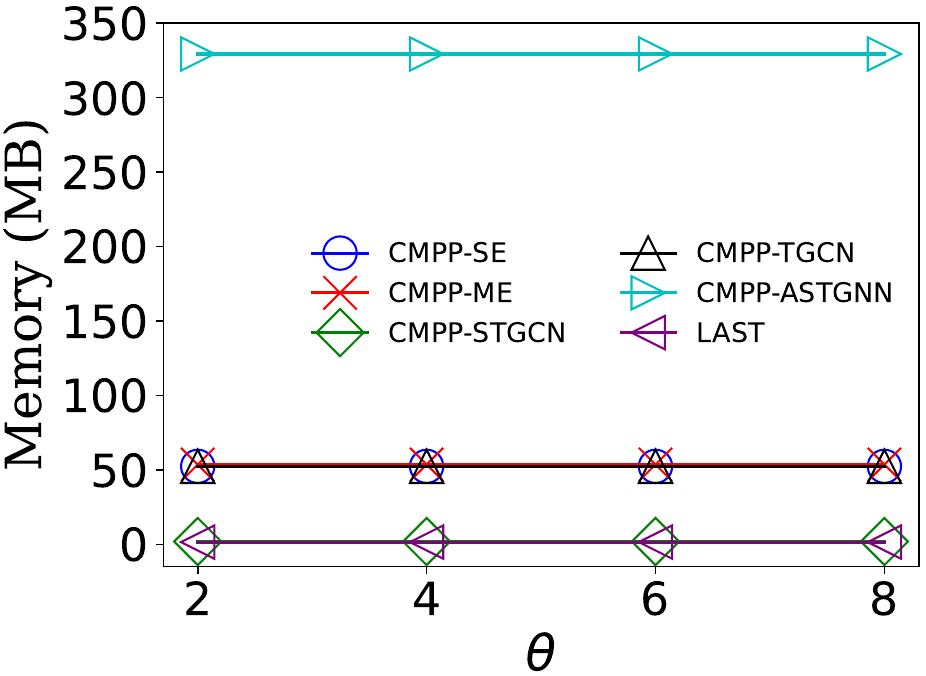}
\ExpCaption{Memory vs $\theta$ (BLD-2).}\label{fig:popu_memo}
\end{minipage}
\centering
\end{figure*}

The query response time, F1-score, and memory usage for BLD-2 are shown in Fig.~\ref{fig:popu_time}, Fig.~\ref{fig:popu_fscore}, and Fig.~\ref{fig:popu_memo} respectively. The trendencies are similar to that in BLD-1 and the explanations can be found in Section~6.3.2 in the paper.

\subsection{Effect of Confidence Threshold $\eta$.}\label{app:effect_eta} 
\begin{figure*}[!ht]
\centering
\begin{minipage}[t]{0.23\textwidth}
\centering
\includegraphics[width=\textwidth]{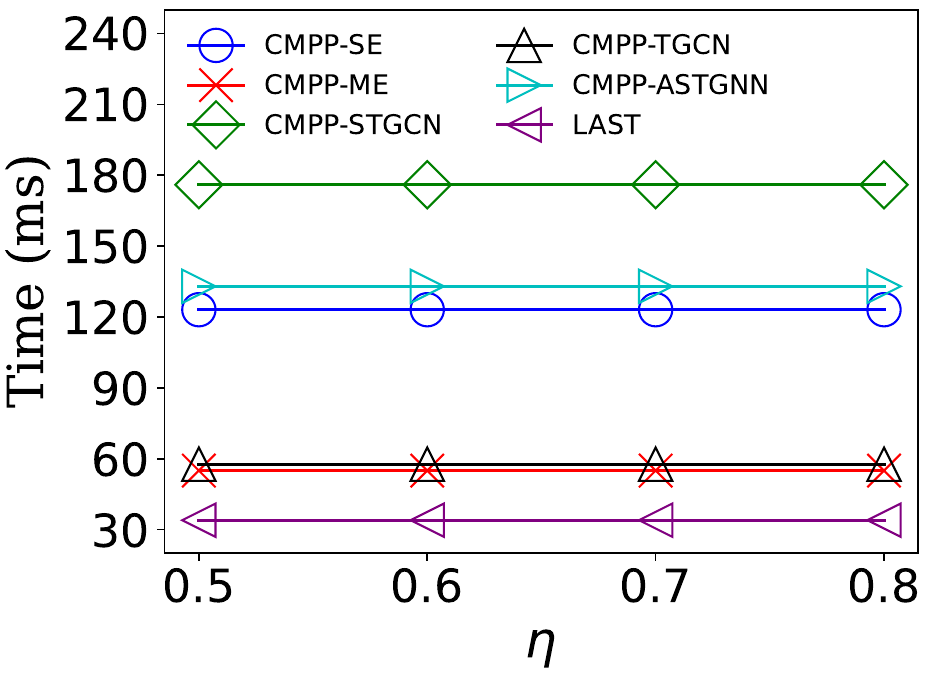}
\ExpCaption{Time vs $\eta$ (BLD-2).}\label{fig:eta_time}
\end{minipage}
\begin{minipage}[t]{0.23\textwidth}
\centering
\includegraphics[width=\textwidth]{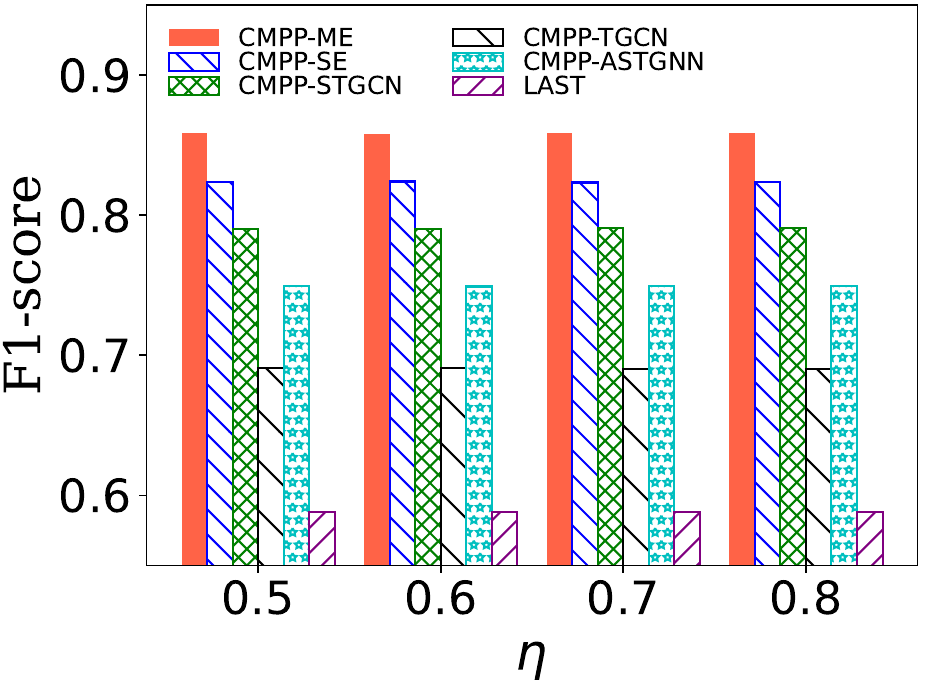}
\ExpCaption{F1-score vs $\eta$ (BLD-2).}\label{fig:eta_fscore}
\end{minipage}
\begin{minipage}[t]{0.23\textwidth}
\centering
\includegraphics[width=\columnwidth]{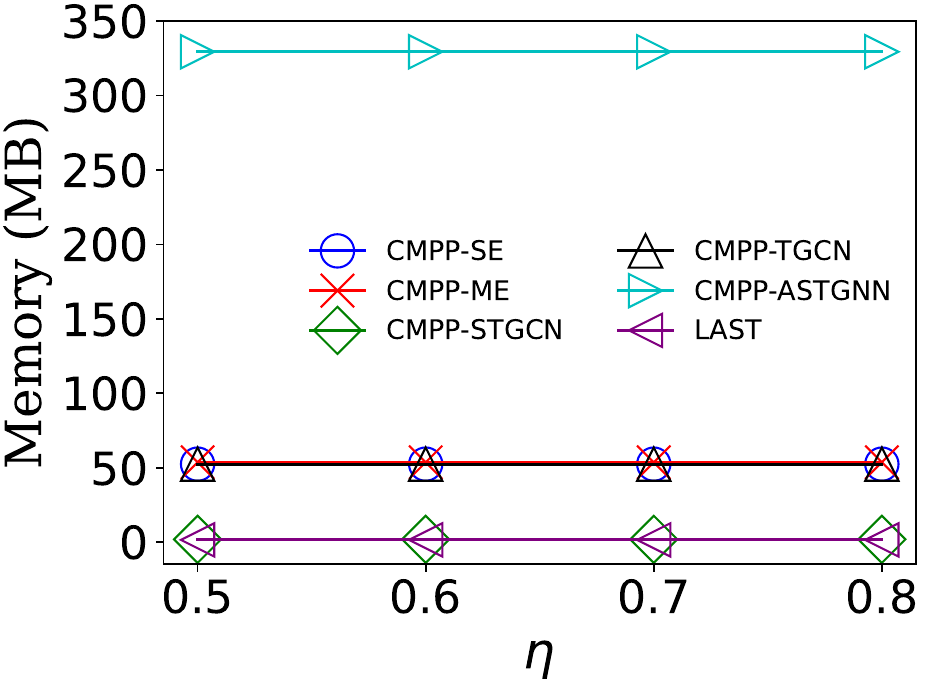}
\ExpCaption{Memory vs $\eta$ (BLD-2).}\label{fig:eta_memo}
\end{minipage}
\centering
\end{figure*}

\begin{figure*}[!ht]
\centering
\begin{minipage}[t]{0.23\textwidth}
\centering
\includegraphics[width=\columnwidth]{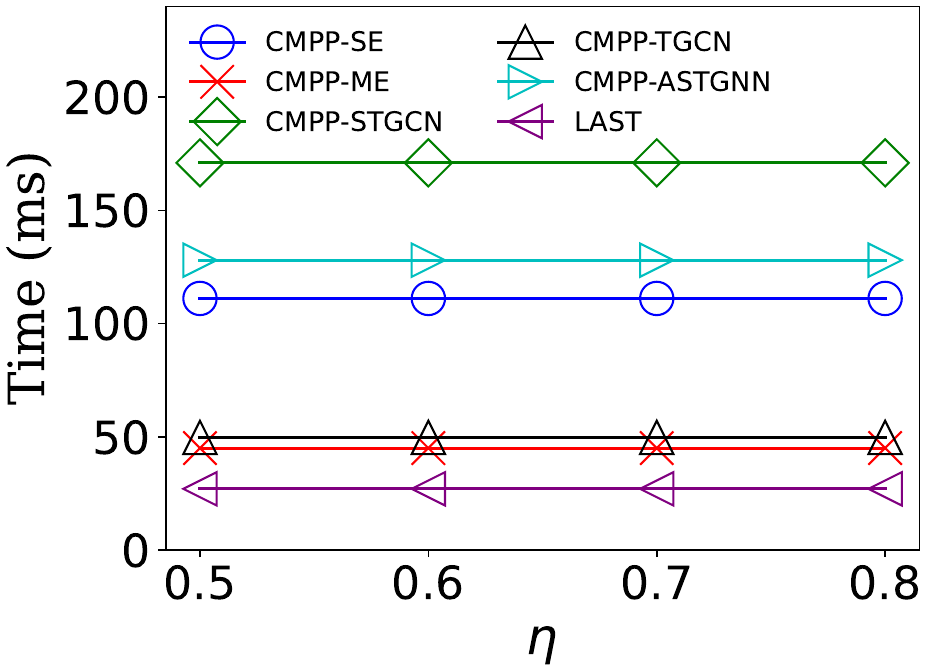}
\ExpCaption{Time vs $\eta$ (BLD-1).}\label{fig:b1_eta_time}
\end{minipage}
\centering
\begin{minipage}[t]{0.23\textwidth}
\centering
\includegraphics[width=\columnwidth]{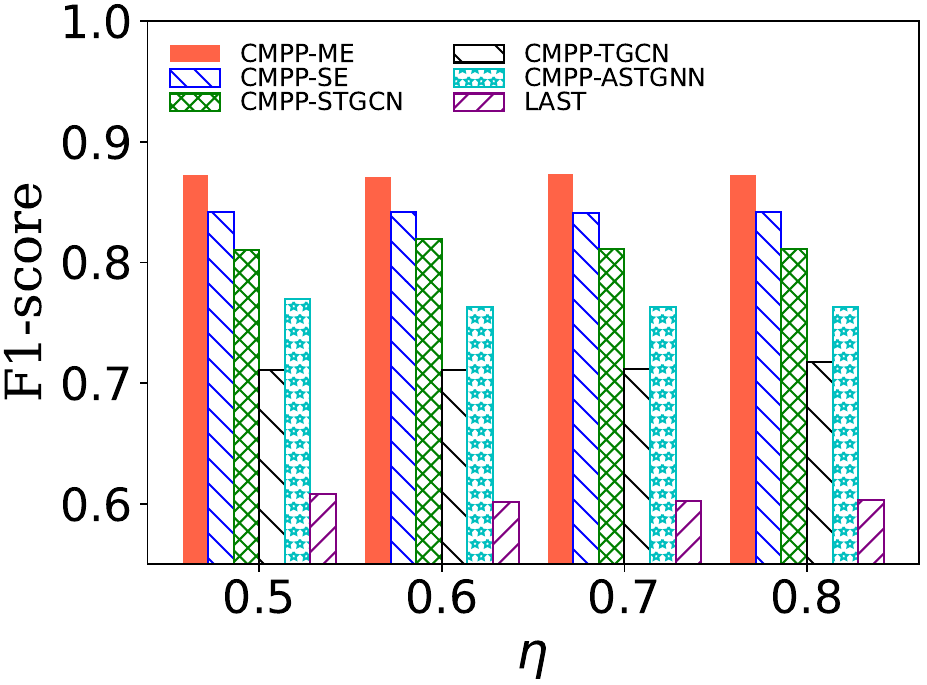}
\ExpCaption{F1-Score vs $\eta$ (BLD-1).}\label{fig:b1_eta_fscore}
\end{minipage}
\begin{minipage}[t]{0.23\textwidth}
\centering
\includegraphics[width=\columnwidth]{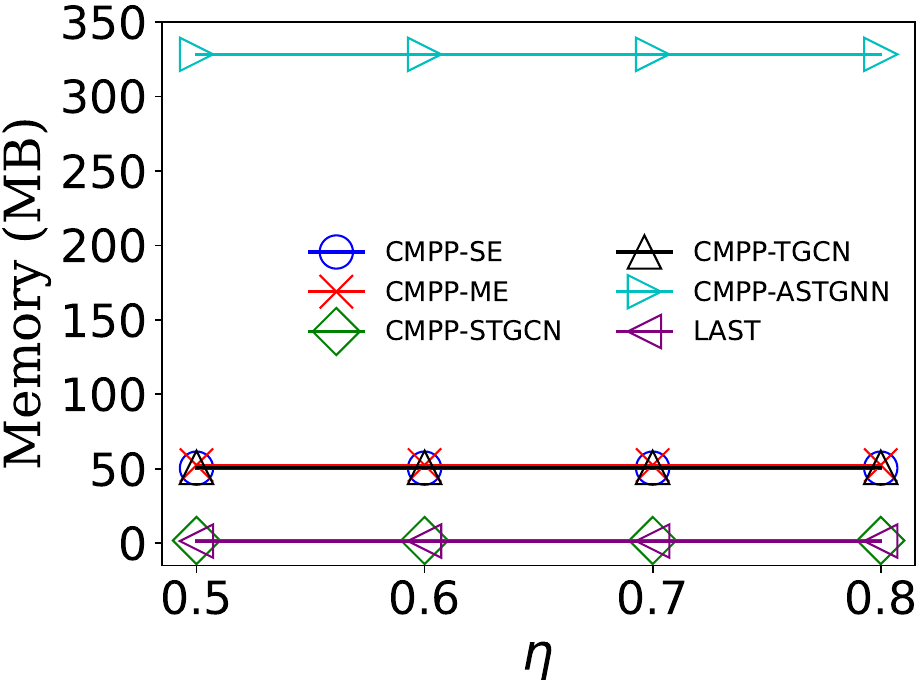}
\ExpCaption{Memory vs $\eta$ (BLD-1).}\label{fig:b1_eta_memo}
\end{minipage}
\centering
\end{figure*}

The query response time, F1-score, and memory usage for BLD-2 are shown in Fig.~\ref{fig:eta_time}, Fig.~\ref{fig:eta_fscore}, and Fig.~\ref{fig:eta_memo} respectively. Besides, the similar results for BLD-1 are complemented and reported in Fig.~\ref{fig:b1_eta_time}, Fig.~\ref{fig:b1_eta_fscore}, and Fig.~\ref{fig:b1_eta_memo} respectively. Similar to $\theta$, $\eta$ as a parameter to define a populated partition has no effect on the query efficiency and effectiveness.

\subsection{Effect of \texttt{Validity}.}\label{app:effect_valid}
\begin{figure*}[!ht]
\centering
\begin{minipage}[t]{0.23\textwidth}
\centering
\includegraphics[width=\textwidth]{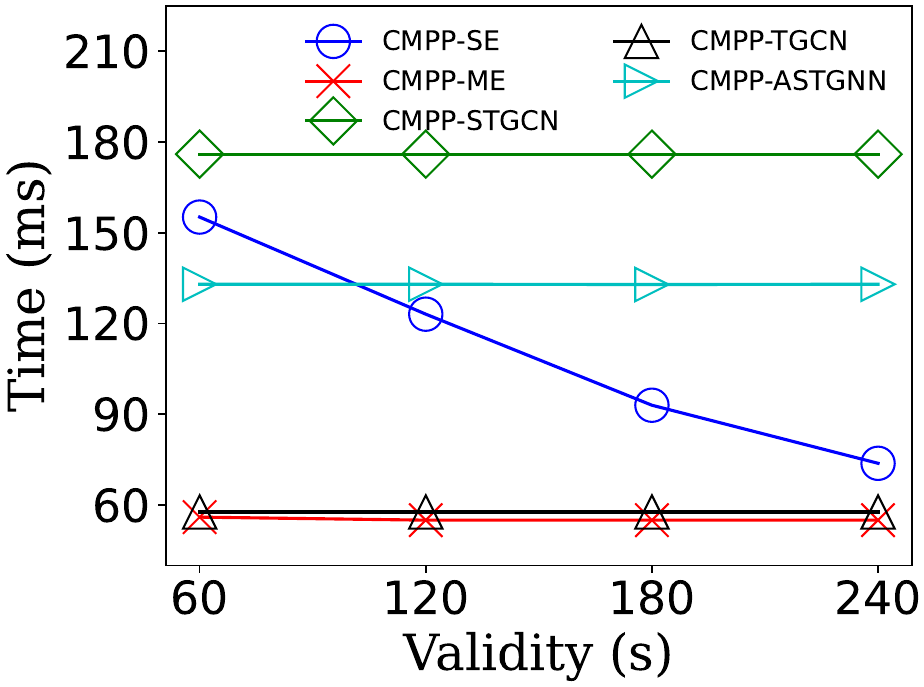}\ExpCaption{Time vs \texttt{Validity} (BLD-2).}\label{fig:duration_time}
\end{minipage}
\begin{minipage}[t]{0.23\textwidth}
\centering
\includegraphics[width=\textwidth]{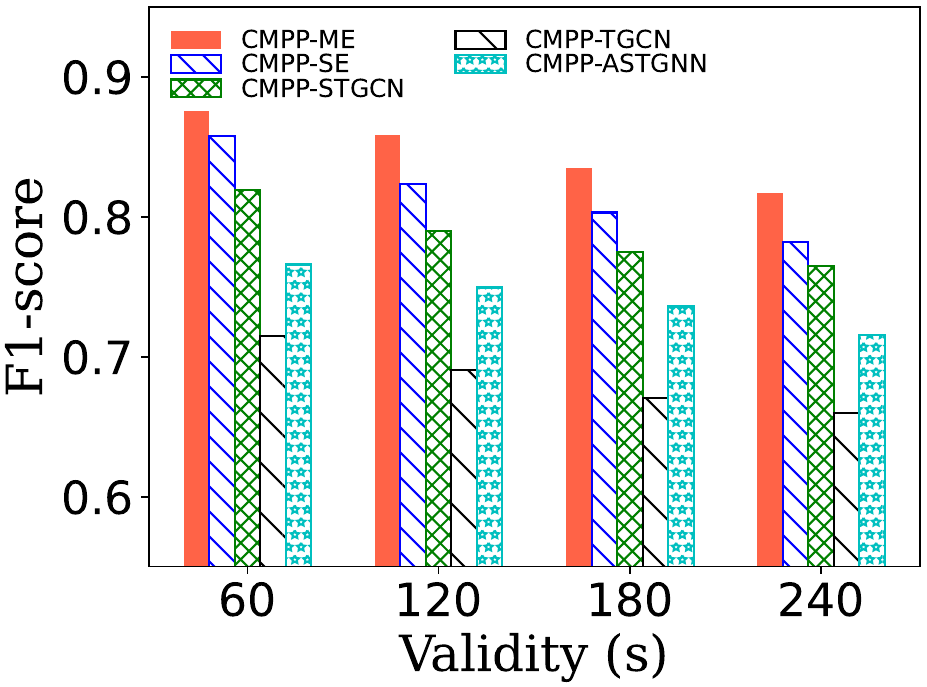}\ExpCaption{F1-score vs \texttt{Validity} (BLD-2).}\label{fig:duration_fscore}
\end{minipage}
\begin{minipage}[t]{0.23\textwidth}
\centering
\includegraphics[width=\columnwidth]{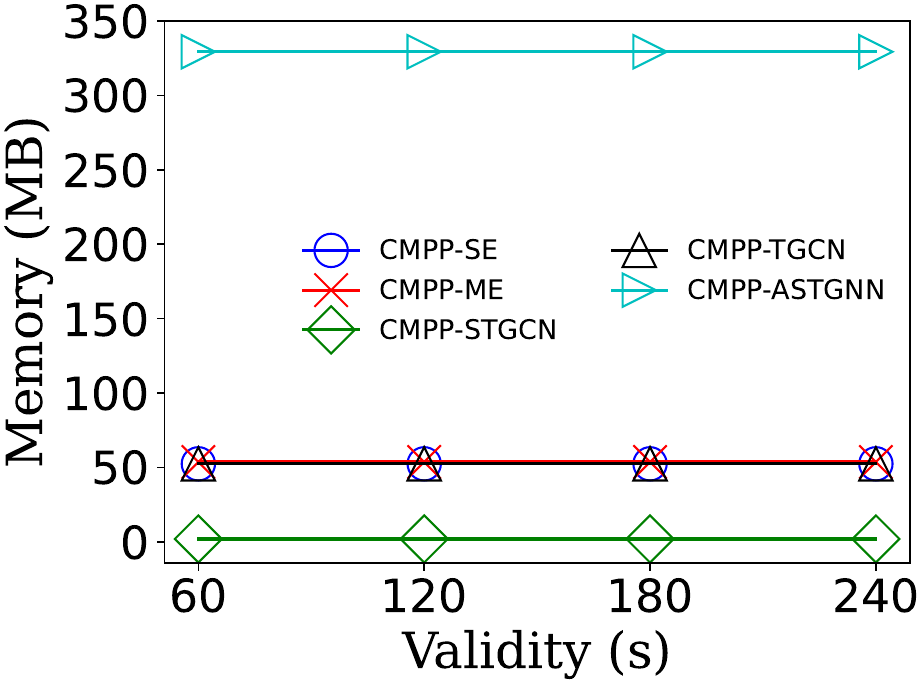}
\ExpCaption{Memory vs \texttt{Validity} (BLD-2).}\label{fig:duration_memo}
\end{minipage}
\centering
\end{figure*}

The query response time, F1-score, and memory usage for BLD-2 are shown in Fig.~\ref{fig:duration_time}, Figure~\ref{fig:duration_fscore}, and Fig.~\ref{fig:duration_memo} respectively. The trendencies are similar to that of BLD-1 and detailed explanations are available in Section 6.3.3 in the paper.

\subsection{Effect of Caching Mechanism.} \label{app:effect_cache}
The query response time and F1-score are reported in Fig.~\ref{fig:snapshot_time} and Fig.~\ref{fig:snapshot_fscore} respectively. Please refer to Section 6.3.4 in the paper for interpretations.

\noindent\textbf{Summary.}
Overall, the results in BLD-2 are similar to that in BLD-1 and this further demonstrates the effectiveness and efficiency of {CMPP} query processing framework. Besides, simlar to $\theta$, varying $\eta$ has almost no effect on the results of query processing. More detailed information are available in Section 6.3 in the paper.

\begin{figure}[H]
\centering
\begin{minipage}[t]{0.23\textwidth}
\centering
\includegraphics[width=\textwidth]{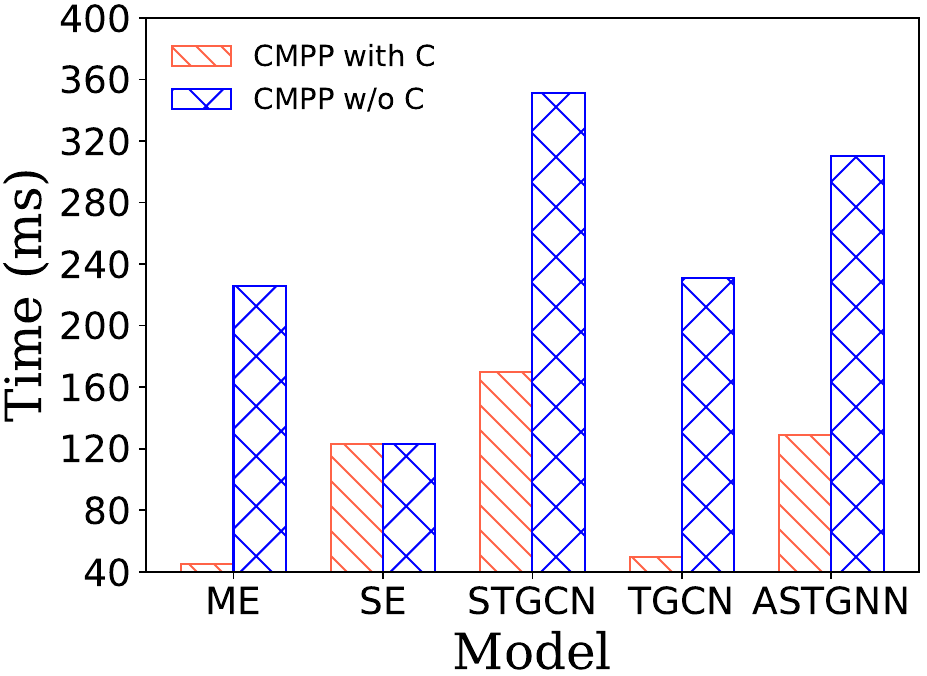}
\ExpCaption{Time vs model (BLD-2).}\label{fig:snapshot_time}
\end{minipage}
\begin{minipage}[t]{0.23\textwidth}
\centering
\includegraphics[width=\textwidth]{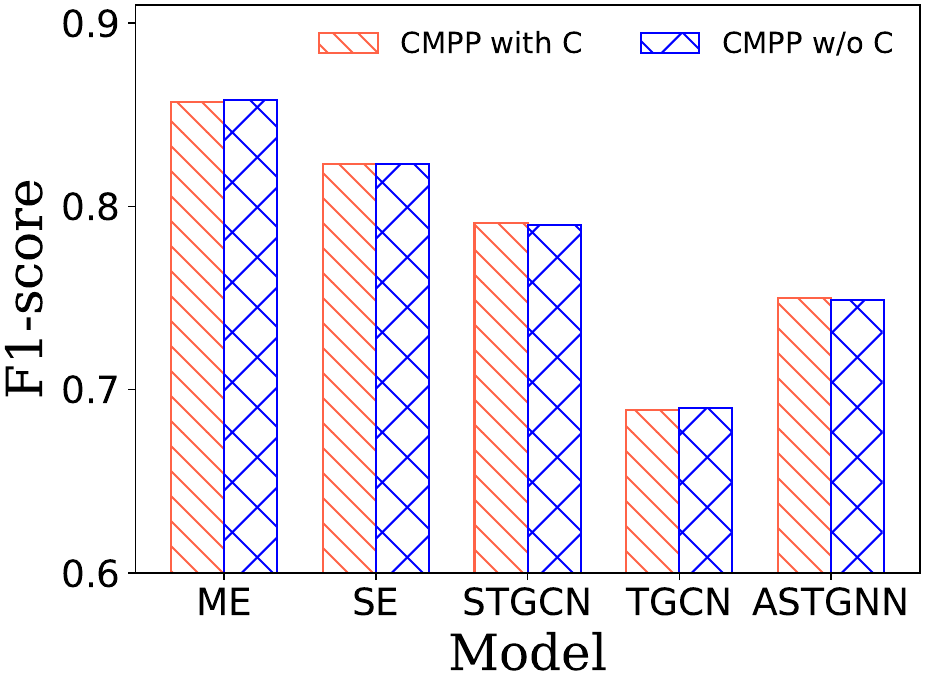}
\ExpCaption{F1-score vs model (BLD-2).}\label{fig:snapshot_fscore}
\end{minipage}
\centering
\end{figure}

\section{Discussion of Loss Functions}
\label{ssec:loss_func}

In this part, we discuss if there are other metrics that can be used as the loss function of our estimators. Specifically,  given the predicted Normal distribution (i.e., $\widehat{P}_{v,t} \sim \mathcal{N}(\hat{\mu}_{v,t}, \hat{\sigma}_{v,t}^2)$) and the ground-truth Normal distribution (i.e., $P_{v,t} \sim \mathcal{N}(\mu_{v,t}, \sigma_{v,t}^2)$), we discuss two alternative loss functions as follows:

\noindent\textbf{Mean Squared Error based Loss.} This is what we initially use as the loss function of our estimators and its specific formula is as follows:
\begin{equation}\label{equation:gru_loss}
\begin{aligned}
\mathcal{L}_1 = 1/2 \cdot (||\mu_{v,t} - \hat{\mu}_{v,t}||_2^2 + \lambda ||\sigma_{v,t}^2 - \hat{\sigma}_{v,t}^2||_2^2),
\end{aligned}
\end{equation}
where $\lambda$ is a weight and set to 1 in our experiments. Given that the co-efficient $1/2$ does not influence the optimization results, we can ignore its effect here. As we can see, $\mathcal{L}_1$ is highly similar to Wasserstein distance-based loss function (i.e., $\mathcal{L}_\text{SE}$ in Equation~\ref{equation:wa_loss})  already introduced before. The only difference between $\mathcal{L}_1$ and $\mathcal{L}_\text{SE}$ is that $\sigma_{v,t}$ here and $\hat{\sigma}_{v,t}$ in $\mathcal{L}_\text{SE}$ (i.e., Equation~\ref{equation:wa_loss})  are replaced with their quadratic forms in $\mathcal{L}_1$. However, when $\hat{\sigma}_{v,t}$ is close to its ground-truth value $\sigma_{v,t}$, $\hat{\sigma}_{v,t}^2$ is close to  $\sigma_{v,t}^2$ as well. Namely, it is very likely that both $\mathcal{L}_1$ and $\mathcal{L}_\text{SE}$ reach the same local optimum point, which explains that the replacement of $\mathcal{L}_1$ with $\mathcal{L}_\text{SE}$ makes little difference to the experimental results. Despite that, the Wasserstein distance-based loss function can steer clear of the hyperparameter of $\lambda$ and increase the interpretability of the loss function, and thus we choose to use Wasserstein distance-based loss function.

\noindent\textbf{KL Divergence based Loss.}
There exist other statistic distance metrics, e.g., Kullback–Leibler (KL) divergence, that could be used to measure the distance between the two Normal distributions. However, KL divergence is featured by non-symmetry and does not satisfy the triangle inequality~\cite{kullback1997information}, which makes it unsuitable for the loss function of our problem.

\end{document}